%% file: main.tex
\documentclass[aps,prx,reprint,superscriptaddress,longbibliography]{revtex4-1}

\usepackage{amsthm,amsmath,amssymb}
\usepackage{tikz}
\usepackage{standalone}
\usepackage{bm}

\usepackage{booktabs}
\usepackage{multirow}
\usepackage{makecell}

 \usepackage{gensymb}

\usepackage{mathrsfs}
\usepackage{graphicx}
\usepackage{amsmath,amssymb,amsfonts,bm}
\usepackage[colorlinks, linkcolor=black, anchorcolor=black, citecolor=blue, runcolor=black, urlcolor=black, CJKbookmarks=true]{hyperref}
\usepackage{lipsum}
\usepackage{enumitem}
\usepackage{array} 
\usepackage{amsmath,amssymb}  
\usepackage{caption}      

\newcommand{\ket}[1]{\left| #1 \right\rangle}

\def\be{\begin{equation}} \def\ee{\end{equation}}
\def\bea{\begin{eqnarray}} \def\eea{\end{eqnarray}}

\def\Tr{\text{Tr}}

\newcommand{\Z}{\mathbb{Z}} 
\renewcommand\[{\begin{equation}}
\renewcommand\]{\end{equation}}

\begin{document}

\title{Generalized symmetry-protected topological phases in mixed states from gauging dualities}

\author{Linhao Li}
\affiliation{Department of Physics, The Pennsylvania State University,University Park,Pennsylvania 16802,USA}

\author{Zhen Bi}
\email{zjb5184@psu.edu}
\affiliation{Department of Physics, The Pennsylvania State University,University Park,Pennsylvania 16802,USA}

\author{Weiguang Cao}
\email{weiguangcao@imada.sdu.dk}
\affiliation{Center for Quantum Mathematics at IMADA, Southern Denmark University, Campusvej 55, 5230 Odense, Denmark}
\affiliation{Niels Bohr International Academy, Niels Bohr Institute, University of Copenhagen, Denmark}

\begin{abstract}  
Decoherence in realistic quantum platforms motivates a mixed-state notion of topological phases of matter, including average symmetry-protected topological (ASPT) phases. Alongside this progress, generalized symmetries—notably noninvertible and dipole symmetries—have become powerful organizing principles for exotic quantum phases, yet their implications for mixed states remain less explored. In this work, we bridge these directions through a gauging correspondence between mixed-state phases with generalized symmetries and mixed-state phases with ordinary group symmetries, recasting the classification of noninvertible and dipole ASPT phases into familiar classifications of symmetry breaking and ASPT phases with dual symmetries. Using this approach, we classify and construct a subclass of ASPT phases with non-invertible and dipole symmetries in $(1+1)d$, including phases that are intrinsic to mixed states, and characterize them via string order parameters and protected edge modes. 

\end{abstract}

\maketitle
\section{Introduction}
Symmetry plays a pivotal role in classifying and understanding phases of matter in quantum many-body systems. In the traditional Landau framework, phases are organized by distinct patterns of spontaneous symmetry breaking (SSB). Over the past decades, however, it has become clear that the Landau paradigm is incomplete: in particular, the discovery of symmetry-protected topological (SPT) phases~\cite{PhysRevB.80.155131,Pollmann:2009ryx,Pollmann:2009mhk,Chen:2011bcp,Chen:2014zvm} has revealed a much richer landscape. Beginning with early examples such as the Haldane chain protected by spin-rotation symmetry~\cite{PhysRevLett.50.1153} and topological insulators protected by charge conservation and time-reversal symmetry~\cite{PhysRevLett.95.226801,PhysRevLett.95.146802,Fu:2006djh,RevModPhys.82.3045,RevModPhys.83.1057}, an extensive variety of SPT phases has since been classified and explored~\cite{Chen:2010zpc,PhysRevB.84.235128,PhysRevB.84.165139,Chen:2011pg,Else:2014vma,Ogata:2020ofz}, forging deep connections across condensed matter physics, high-energy theory, and quantum information science. Recent progress has further broadened this landscape along two frontiers: extending SPT physics beyond ordinary group symmetries~\cite{Thorngren:2019iar}, and extending it beyond closed, pure-state settings to mixed states and open quantum systems~\cite{de2022symmetry,PhysRevX.13.031016, zhang2022strange}.

Generalized symmetries have emerged as a powerful language for organizing exotic quantum matter. Two extensions are particularly prominent. The first relaxes the usual group notion of symmetry to algebraic structures that need not be invertible—so-called noninvertible symmetries—whose fusion and associativity data are naturally encoded by fusion categories rather than groups~\cite{Gaiotto:2014kfa,Bhardwaj:2017xup,Tachikawa:2017gyf,Chang:2018iay,Thorngren:2021yso}. The second concerns dipole, or more broadly spatially modulated, symmetries~\cite{Gromov:2018nbv,Gorantla:2022eem,Lake:2022ico,Lake:2022hel,Gorantla:2022ssr,Lam:2024smz,yan2024generalizedkramerswanierdualitybilinear,10.21468/SciPostPhys.17.4.104}, which intertwine internal transformations with spatial structure and can impose kinetic constraints beyond conventional on-site symmetry. Their consequences have been explored in both continuum quantum field theory and microscopic lattice models~\cite{Bhardwaj:2023fca,Bhardwaj:2024wlr,Chen:2025uno}. Together, these generalized symmetries support new families of SPT phases~\cite{Choi:2024rjm,Jia:2024bng,Li:2024fhy,Jia:2024zdp,Jia:2024wnu,Meng:2024nxx,Li:2024gwx,Aksoy:2025rmg,Han:2023fas,Lam:2023xng,Saito:2025qrp, doi:10.1142/S0217751X25480021}, including phases that go beyond the conventional group-cohomology framework~\cite{Seifnashri:2024dsd,Inamura:2024jke,Cao:2025qhg}. Yet much of this progress has focused on closed systems and pure states, leaving it largely open how generalized symmetries constrain open many-body systems and mixed-state phases.

On the other hand, there has recently been a surge of interest in \emph{mixed-state} topological phases in \emph{open} many-body systems subject to decoherence and disorder~\cite{PhysRevX.14.031044,PhysRevLett.134.070403,yang2025topological,sang2025mixed,PRXQuantum.6.010344,gu2024spontaneous,PRXQuantum.6.010314,PRXQuantum.6.010315,PRXQuantum.6.010313,luo2025topological}. In these settings the system is naturally described by a density matrix $\rho=\sum_{I} P_{I}\,|\Psi_I\rangle\langle\Psi_I|$, arising either from coupling to an environment or from statistical mixing~\cite{PhysRevLett.43.1434,PhysRevB.22.1305,PhysRevLett.48.344,PhysRevLett.74.1226,PhysRevLett.103.047201,preskill2018quantum,PhysRevLett.126.130403,PhysRevLett.127.270503,PRXQuantum.3.040313}. The notion of symmetry is correspondingly enriched and naturally splits into two types: a \emph{strong} (or exact) symmetry $K$ and a \emph{weak} (or average) symmetry $G$. Concretely, all components $|\Psi_I\rangle$ carry the same $K$-charge (so $\rho$ lies within a fixed $K$-symmetry sector), while their $G$-charges may differ (so $\rho$ mixes different $G$-sectors). Equivalently, $\rho$ may be viewed as a \emph{canonical} ensemble with respect to $K$, but a \emph{grand-canonical} ensemble with respect to $G$. Within this framework, \emph{average symmetry-protected topological} (ASPT) phases extend the familiar notion of pure-state SPTs to mixed states~\cite{Lee_2025,PhysRevX.13.031016,ma2025topological,PhysRevX.15.021060,xue2024tensor,you2024intrinsic,PRXQuantum.6.020333,PRXQuantum.6.010347}. ASPT phases can exhibit nontrivial edge structure and topological responses protected jointly by strong and weak symmetries. Systematic classifications are developed for ASPTs protected by ordinary group symmetries, in both decohered and disordered settings~\cite{PhysRevX.13.031016}. Moreover, recent works have identified phases that are \emph{intrinsic average}: they admit no pure-state counterpart with the same symmetry and topological response. This perspective highlights that controlled noise and disorder can stabilize genuinely new topological phenomena beyond those accessible in closed, pure-state systems.

A natural question is how these two developments---generalized symmetries on the one hand, and the strong/weak symmetry structure of mixed states on the other---interact, especially in the context of SPT phases. Recent works have begun to explore mixed-state phenomena using the symmetry topological field theory (SymTFT) framework, which provides a systematic description of generalized symmetries. These studies, however, have mainly addressed noninvertible symmetry breaking and mixed-state phases with conventional group symmetries using the doubled Hilbert space technique~\cite{luo2025topological,Schafer-Nameki:2025fiy,Qi:2025tal}, anomalies of weak noninvertible symmetries in the disordered systems \cite{li2026average} and mixed-state phases obtained from the higher-order subsystem SPT phases with noninvertible symmetries \cite{mana2026mixedstatephaseshigherordersspts}. A direct understanding of \emph{mixed-state phases with generalized symmetries}---formulated in terms of two-way channel connectivity of density matrices---is lacking. This gap motivates several concrete questions: Can mixed-state phases with noninvertible or dipole symmetries, in particular ASPT phases, be classified within the two-way connectivity framework? Can we construct representative lattice models, and determine which phases are inherited from pure-state SPTs and which are \emph{intrinsic average} (IASPT), i.e., unique to the mixed-state setting?

\begin{figure}[t]
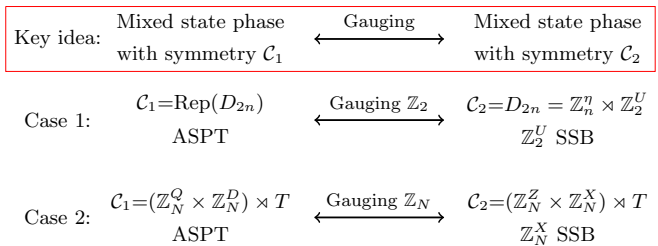

    \centering
\includestandalone[width=1.0\linewidth]{duality}
    \caption{Illustration of gauging-based method to study ASPT phases from group SSB phases. The strong and weak conditions of symmetry operators are shown in \eqref{eq: symmetry conditions-1} and \eqref{eq: symmetry conditions-2}. 
    }
    \label{fig:dualityfigure}
\end{figure} 
In this work, we develop a gauging-based framework to classify broad classes of noninvertible and dipole symmetries in mixed-state settings, building on gauging methods originally formulated for pure-state SPTs~\cite{PhysRevB.86.115109,PhysRevLett.112.141602,Cao:2025qhg}. We view gauging as a concrete quantum operation that relates mixed states with generalized symmetries, denoted $\mathcal{C}_1$, to mixed states with conventional group symmetries, denoted $\mathcal{C}_2$, as shown in Fig. \ref{fig:dualityfigure}. In this language, we demonstrate that gauging induces a one-to-one correspondence between mixed-state phases protected by $\mathcal{C}_1$ and those protected by $\mathcal{C}_2$, where phases are defined via two-way connectivity under symmetry-preserving finite depth local quantum channels. In particular, the classification of ASPT phases protected by noninvertible or dipole symmetries is reduced to familiar group-theoretic classifications of symmetry breaking and SPT phases with the corresponding dual group symmetries. Operationally, we start from mixed-state phases with ordinary internal symmetries and apply appropriate gauging procedures that generate: (i) noninvertible symmetries by gauging non-normal subgroups, and (ii) dipole symmetries by gauging symmetries with Lieb--Schultz--Mattis (LSM)-type anomalies~\cite{Lieb:1961aa,Affleck:1986aa,OYA1997,Oshikawa:2000aa,Hastings:2004ab,Chen-Gu-Wen_classification2010,Fuji-SymmetryProtection-PRB2016,Watanabe:2015aa,Ogata:2018aa,Ogata:2020aa,Yao:2021aa,Yao:2019aa,PhysRevB.110.045118,PhysRevB.106.224420,PhysRevLett.133.136705,PhysRevB.110.045118,PhysRevB.106.224420}. 
Beyond classification, we provide explicit lattice realizations, analyze the resulting protected edge structure, and develop string-order diagnostics. This also clarifies when the resulting generalized-symmetry ASPT phases are intrinsic to the mixed-state regime. Finally, because these gauging procedures can be implemented by finite-depth circuits with ancillas and measurements~\cite{PhysRevX.14.021040}, our framework yields a constructive preparation scheme for mixed states realizing ASPT order with noninvertible or dipole symmetries.


\section{Preliminaries\label{sec:preliminaries}}
\subsection{Group symmetries in open quantum systems\label{sec:sym}}
We first review the strong and weak group symmetry of mixed states and their action on the evolution of open systems generally described by quantum channels. Consider bosonic systems with internal symmetry of group $\Gamma$.  

A density matrix $\rho$ has the $\Gamma$ symmetry iff:
\begin{equation}
\begin{aligned}
 U(g)\rho =\rho U(g),\forall g\in G& \\
\text{ (weak symmetry condition)}&,\\
 U(k)\rho= \lambda(k)\rho,\forall k \in K & \\
\text{ (strong symmetry condition)}&, 
\end{aligned}
\end{equation}
where  $U(g)$ and $U(k)$ are the unitary transformations of weak and strong symmetry respectively, $\lambda(k)$ is a $U(1)$ phase factor depending on the charge carried by $\rho$ under the strong symmetry $K$. Here $K(G)$ is the strong (weak) symmetry group. $K$ must be a normal subgroup of $\Gamma$ and $G$ is the corresponding quotient group $G=\Gamma/K$. 

A general quantum channel can be constructed through the following steps according to the Stinespring dilation theorem: $(i)$ introduce an auxiliary environment degree of freedom enlarging the Hilbert space $\mathcal{H}^S$ to $\mathcal{H}^S\otimes \mathcal{H}^E$ with the environment initialized in a pure state $|0\rangle_E$. $(ii)$ Apply a unitary evolution $\mathcal{U}$ to the total system $S\cup E$. $(iii)$ Trace out the environment. The resulting quantum channel $\mathcal{N}$ takes the form
\begin{equation}
\mathcal{N}[\rho]=\Tr_E\{\mathcal{U}\rho\otimes|0\rangle_E\langle 0|_E \mathcal{U}^\dagger\}.
\label{eq:dilation}
\end{equation}
Alternatively, a general quantum channel can also be expressed in terms of Kraus operators:
\begin{equation}
\mathcal{N}[\rho]=\sum_i K_i \rho K^{\dagger}_i, \quad \text{with}\ \  \sum_i K^{\dagger}_iK_i=I.
\end{equation}
In connection with the dilation form, each Kraus operator can be written as $K_i=\langle b_i|\mathcal{U}|0\rangle_E$ where $|b_i\rangle$ is a basis of the environment Hilbert space. A quantum channel is symmetric under $\Gamma$ symmetry iff
\begin{equation}
\begin{aligned}
&\mathcal{N} \text{ is invariant under }
K_i\rightarrow U(g)K_i U^\dagger(g),\forall g\in G \\& \qquad\qquad\qquad\qquad\qquad\text{ (weak symmetry condition)},\\
&[U(k),K_i]=0,\forall k\in K \\
& \qquad\qquad\qquad\qquad\qquad\text{ (strong symmetry condition)}.
\end{aligned}
\end{equation}


For the purpose of classifying phases of matter, we focus on quantum channels that are \emph{local} and \emph{finite-depth}, which capture the physically relevant notion of finite time local evolution in open quantum systems. Following Ref.~\cite{hastings2011topological}, a finite-depth local channel (FDLC) requires that environment Hilbert space factorize as $\mathcal{H}^E=\otimes_i \mathcal{H}^E_i$ on each site with the environment initialized in a product state $|0\rangle_E$ and the joint unitary acting on the system and its environment is chosen to be a finite-depth local unitary (FDLU) circuit. The FDLC provides a natural generalization of FDLU for pure states. 

In discussing symmetry-protected phases of matter, we restrict attention to quantum channels that respect the symmetry \emph{locally} \cite{de2022symmetry, PhysRevX.13.031016,ma2025topological}. We say that an FDLC is \emph{$\Gamma$-locally symmetric} if it admits a purification such that the joint system--environment unitary can be written as an FDLU and satisfy 
\begin{equation}
\begin{aligned}
[U(g)\otimes U_E(g),\mathcal{U}^{(j)}]&=0,U_E(g)|0\rangle_E=\lambda_E(g)|0\rangle_E,\forall g\in G\\
&\text{(weak symmetry condition)},\\
[U(k)\otimes I_E,\mathcal{U}^{(j)}]&=0,\forall k\in K & \\
&\text{(strong symmetry condition)},
\end{aligned}
\end{equation}
where $\mathcal{U}^{(j)}$ represents local gates of the unitary circuit. Here $U_E(g)$ is a unitary representation of $G$ acting on the environment, and $\lambda_E(g)\in U(1)$ is a phase factor defining a one-dimensional representation of $G$ on the reference environment state $|0\rangle_E$ \footnote{For non-onsite symmetries, we need each layer of local unitary gates to be invariant under the symmetry action.}.




\subsection{Mixed-state phases}
When defining  phases of matter, one needs an equivalence relation between phases. We review the recently proposed definitions of mixed-state quantum phases. We begin with phases in the absence of symmetry.

\textbf{Definition 1} (mixed-state quantum phases). Two mixed states $\rho_1$, $\rho_2$ belongs to the same phase iff they are two-way connected by FDLCs, that is, $\exists$ a pair FDLCs $\mathcal{N}_{12},\mathcal{N}_{21}$, s.t. $\rho_2=\mathcal{N}_{12}[\rho_1],\rho_1=\mathcal{N}_{21}[\rho_2]$. 

This definition is a natural generalization of  the classification of ground-state quantum phases of local gapped Hamiltonians under equivalence by FDLUC \cite{PhysRevB.82.155138,PhysRevB.83.035107}.  A subtlety arises because FDLCs are generally not invertible, which necessitates requiring two-way connectivity. Similar to the pure-state setting, symmetries can further enrich the mixed-state phase structure:

\textbf{Definition 2} (Symmetric mixed-state quantum phases). Two mixed states with the $\Gamma$ symmetry belong to the same symmetric phase iff they are two-way connected by $\Gamma$ locally-symmetric FDLCs.

In particular, the nontrivial ASPT refers to the states that can be two-way connected to a product state only if the FDLC is not locally-symmetric.  The general classification of ASPT with group symmetries has been systematically investigated in \cite{PhysRevX.13.031016,ma2025topological}.

Here we remark that Ref.~\cite{sang2025mixed} recently proposed a refined notion of mixed-state phase equivalence based on two-way connectivity via \emph{locally reversible} channel circuits. This refinement yields a finer classification of mixed-state phases, notably distinguishing classical memory as a phase that is inequivalent to the trivial state. Since equivalence under the refined definition implies equivalence under the original two-way connectivity criterion~\cite{PhysRevLett.134.070403}, any two states that are \emph{not} two-way connected must necessarily belong to distinct mixed-state phases within the refined framework. For the rest of our results, we adopt the coarser two-way connectivity condition which provides a correspondingly coarse classification of ASPT phases. However, based on previous understanding~\cite{ma2025topological}, the original two-way connectivity criterion appears to capture all robust topological data of ASPT phases. At present, we are not aware of examples in which the refined definition yields additional, genuinely new phases in the ASPT setting.

\subsection{Noninvertible symmetry from duality}
In this subsection we review an important class of noninvertible symmetries that can be obtained from group symmetries by duality transformations. 
It is known that gauging a non-normal subgroup of a group symmetry results in a noninvertible symmetry in the dual model~\cite{Bhardwaj:2022kot,Bhardwaj:2022maz,Hsin:2024aqb,Hsin:2025ria,Cao:2025qnc}. 
In this work, we focus on the examples obtained from gauging the non-Abelian 0-form $D_{2m}$ symmetry~\footnote{The superscript $(0)$ for 0-form symmetry is omitted when unambiguous. Instead, we use the superscript for the generator of the group.} 
and their lattice constructions in $(1+1)d$, which is relevant for the later study of average noninvertible SPT (ANISPT).  

We start with the non-abelian 0-form 
$D_{2m}=\mathbb Z_m^{\eta}\rtimes \mathbb Z_2^{U}$ symmetry, generated by $\mathbb Z_m$ operator $\eta$ and $\mathbb Z_2$ operator $U$, in $(1+1)d$. 
We get the dual noninvertible symmetry by gauging   $\mathbb Z_2^{U}$, which is equivalent to the Kramers-Wannier (KW) duality~\cite{Kramers:1941kn,Frohlich_2004}, and the dual symmetry is Rep$(D_{2m})$ symmetry~\cite{Decoppet:2024htz,Choi:2024rjm,Bhardwaj:2022maz,Cao:2025qhg}.  The symmetry operators correspond to the irreducible representations of the  group $D_{2m}$.   Notably, the noninvertible operators are constructed from $\eta^l+\eta^{-l}$, the $\Z^U_2$-even combination of $\Z_m$ operators.
After gauging, we get noninvertible symmetry operators $\mathsf{W}_l$ with the following fusion rule
\begin{equation}
\mathsf{W}_l\times \mathsf{W}_k=\mathsf{W}_{l+k}+\mathsf{W}_{l-k},
\quad l,k\in \Z_m\,,
\end{equation}
which directly follows from the fusion rule of these $\Z^U_2$-even combinations. For $m=0$~mod~$2$, $\eta^{m/2}$ gives rise to a $\mathbb{Z}_2$ symmetry operator.
In addition, gauging also gives rise to a quantum $\mathbb{Z}_2$ symmetry. 

In preparation for the lattice realization of ANISPTs, we review the construction of Rep$(D_{2m})$ symmetry in the $(1+1)d$ spin chain with $L$ sites and a tensor product Hilbert space $(\mathbb C^2)^L$. We focus on the case with even $m$, i.e., $m=2n$. For convenience, we will use $D_{4n}$ instead.
Starting from a model with  $D_{4n}$ symmetry generated by 
\begin{equation}\label{eq: symmetry action}
 \eta=\prod_{j=1}^L\exp\left(\frac{\pi i}{2n}(1-\sigma^z_j)\right), \quad U=\prod_{j=1}^{L}\sigma^x_j,
\end{equation}
we perform the KW duality transformation, equivalent to gauging $\mathbb Z_2^{U}$ symmetry
\begin{equation}\label{eq:kw1dlattice}
\mathcal{D}: \sigma^x_{j}\to \sigma^z_{j-1}\sigma^z_{j},\quad \sigma^z_{j}\sigma^z_{j+1}\to \sigma^x_{j},
\end{equation}
and obtain the dual model with Rep$(D_{4n})$ symmetry.  The objects of  Rep$(D_{4n})$ are irreducible representations (irreps) of $D_{4n}$ and their fusion algebra follows from the tensor product of these irreps.  The dual Rep$(D_{4n})$ symmetry is generated by
\begin{align}
    \begin{split}
        &U_e=\prod_{k=1}^{L/2}\sigma^x_{2k},\quad U_o=\prod_{k=1}^{L/2}\sigma^x_{2k-1},\\
        &\mathsf{W}_{l}=\mathcal{D}^{\dagger}(\eta^l+\eta^{-l})\mathcal{D},\quad l=1,...,n-1,
    \end{split}
\end{align}
where $U_e,U_o$ generate the invertible $\mathbb Z^e_2\times \mathbb Z^o_2$ spin-flip symmetry on even and odd sites and $\mathsf{W}_{l}$ generates the noninvertible part. These noninvertible operators are also given by matrix product operators (MPO) with 2-dimensional virtual bond~\cite{Cao:2025qnc}:
\[
\begin{split}
&\mathsf{W}_{l}=\text{Tr}\left(\delta_{\alpha_1,\alpha_{L+1}}\prod^L_{j=1}M^{(j)}_{\alpha_j\alpha_{j+1}}\right),\\
&M^{(j)}=\begin{bmatrix}
    \exp\left(\frac{i\pi  l}{n}\right)|+\rangle\langle+|_j       & \exp\left(\frac{-i\pi  l}{n}\right)|-\rangle\langle-|_j  \\
    \exp\left(\frac{i\pi  l}{n}\right)|-\rangle\langle-|_j       &\exp\left(\frac{-i\pi  l}{n}\right)|+\rangle\langle+|_j
\end{bmatrix},
\end{split}
\]
where $|+\rangle_j$ and $|-\rangle_j$ are eigenvectors of $\sigma^x_j$ with eigenvalue $\pm 1$ respectively. 

For example, when $n=2$ we have Rep($D_8$) symmetry. Besides a $\mathbb Z^e_2\times \mathbb Z^o_2$ subgroup generated by two invertible symmetry operators $U_e$ and $U_o$, there is one noninvertible symmetry operator $\mathsf{W}_{1}$ with fusion rule
\begin{equation}
    \mathsf{W}_{1}\times \mathsf{W}_{1} =1+U_e+U_o+U_eU_o.
\end{equation}
$\mathsf{W}_{1}$ also induce a mapping on $\mathbb Z^e_2\times \mathbb Z^o_2$ symmetric local operators as follows \cite{PhysRevB.108.214429}:
\begin{equation}\label{eq: W1 trans}
    \mathsf{W}_{1}: \sigma^x_j \to\sigma^x_j,\quad  \sigma^z_{j-1} \sigma^z_{j+1}\to -\sigma^z_{j-1}\sigma^x_j\sigma^z_{j+1}.
\end{equation}

\subsection{Dipole symmetry from duality}\label{sec:dipolesym}
Modulated symmetries, especially dipole symmetries, can emerge by gauging a subgroup of the internal symmetry sharing the LSM-type anomaly with lattice translation symmetry~\cite{Aksoy:2023hve,Ebisu:2025mtb}. 
Let's focus on the simplest example in $(1+1)d$ with $\mathbb Z_N$ spin on a chain with length $L\in N\Z$, which is relevant for the discussion of dipole ASPT later. Consider internal $\mathbb Z_N^X\times \mathbb Z_N^Z$ symmetry generated by 
\begin{equation}\label{eq:LSM}
    U_X=\prod_{j=1}^LX_j,\quad U_Z=\prod_{j=1}^LZ_j.
\end{equation}
Since $X_j$ and $Z_j$ on each site exhibit a nontrivial commutation relation with an anomalous phase, e.g., $X_j Z_j=\omega Z_j X_j$ with $\omega=\exp(2\pi i/N)$, there is an LSM anomaly between internal symmetry and lattice translation symmetry. If we gauge $\mathbb Z_N^X$ symmetry, equivalent to the $\mathbb Z_N$ KW transformation
\begin{equation}\label{eq:ZNKW}
    \mathsf{D}:Z_{j-1}^{\dagger}Z_{j}\to X_{j-1},\quad  X_{j}\to Z_{j-1}^{\dagger}Z_{j},
\end{equation}
we get the dual theory
with $\mathbb Z_N^Q\times \mathbb Z_N^D$ charge and dipole symmetry generated by
\begin{equation}\label{eq:dipolesym}
    U_Q=\prod_{j=1}^LX_j,\quad U_D=\prod_{j=1}^L(X_j)^j,
\end{equation}
where $U_Q$ generates the dual $\mathbb Z_N^Q$ symmetry and $U_D$ is the image of $U_Z$ after the duality transformation~\eqref{eq:ZNKW}.


Because of the KW duality~\eqref{eq:ZNKW}, we can classify phases with dipole symmetry~\eqref{eq:dipolesym} in both closed and open quantum systems, from the knowledge of phases with $\mathbb Z_N^X\times Z_N^Z$ symmetry~\eqref{eq:LSM} with LSM anomaly.

\section{Classification of $(1+1)d$ \text{Rep}($D_{2m})$ average SPT}
In this section, we will identify Rep($D_{2m}$) symmetry in open quantum systems and then classify the associated ASPT in $(1+1)d$ using the gauging method. We emphasize that the gauging procedure can only be done with strong symmetries of the system. 

\subsection{\text{Rep}($D_{2m})$ symmetry in open quantum systems}\label{sec:repd2mopen}
Now we consider the $(1+1)d$ mixed states in open quantum systems. The strong symmetry can only be a normal subgroup of 
$D_{2m}=\mathbb Z_m^{\eta}\rtimes \mathbb Z_2^{U}$. Because the dual models with non-invertible symmetries are obtained by gauging the $\mathbb Z_2^U$ subgroup, we require $\mathbb Z_2^U$ to be a strong symmetry. Thus, we need $\mathbb Z_2^U$ to be part of the normal subgroup that forms a consistent strong symmetry of the system. This leads to two possible scenarios of the strong symmetry depending on the parity of $m$
\begin{enumerate}
\item  If $m$ is odd, the normal subgroup, i.e., the strong symmetry group, can only be the whole $D_{2m}$ group.
\item If $m$ is even, there are two choices of strong symmetry $K$. Besides the whole $D_{2m}$ group, the other normal subgroup is $K=\mathbb Z_{m/2}^{\eta^2}\rtimes\mathbb Z_2^{U}=\{\eta^2|\eta^m=1\}\rtimes \{1,U\}$ which results in a weak symmetry group $G=\Z_2$. 
\end{enumerate}
 If the strong symmetry is $D_{2m}$, then the dual $\text{Rep}(D_{2m})$ symmetry after gauging will be strong, where the classification of ASPT should be the same as pure state cases. Hence we will focus on the second case with even $m$ with the convention $m=2n$, and we have the mixed strong-weak noninvertible Rep($D_{4n}$) symmetry, where the noninvertible symmetry contains both strong and weak parts.  In such a case, $\mathsf{W}_{2k-1}$ is the weak symmetry and the other operators are all strong symmetry:
\begin{equation}\label{eq: symmetry conditions-1}
    \mathsf{W}_{2k-1} \rho=\rho \mathsf{W}_{2k-1}, \, \mathsf{W}_{2k}\rho =\lambda_{2k} \rho, \, U_{e/o}\rho=\lambda_{e/o}\rho.
\end{equation}
Here $\lambda_{2k}$ is a complex number whose norm is the quantum dimension of $\mathsf{W}_{2k}$. This is a direct generalization of the strong/weak symmetry condition of group symmetry to noninvertible Rep($D_{4n}$) symmetry.

\subsection{Gauging as a one-to-one mapping between mixed-state phases}\label{sec: gauging map}
For pure states, gauging establishes a one-to-one correspondence between quantum phases before and after gauging. We show here this principle extends naturally to mixed states as well, and it will play a fundamental role in our classification.

 For concreteness, let us consider gauging $\Z^U_2$ symmetry, a duality between $D_{4n}$ and $\text{Rep}(D_{4n})$, although this specific choice is not essential and the argument below can be applied to other dual symmetries. We begin with FDLC $\mathcal{N}$ which locally preserves $D_{4n}$. That is  
\[
\begin{split}
&\mathcal{N}[\rho]=\Tr_E\{\mathcal{U}\rho\otimes|0\rangle_E\langle 0|_E \mathcal{U}^\dagger\},\\
&[U\otimes I_E,\mathcal{U}^{(j)} ]=[\eta^2\otimes I_E,\mathcal{U}^{(j)} ]=[\eta\otimes \eta_{E},\mathcal{U}^{(j)}]=0,\nonumber
\end{split}
\]
where $\eta_E$ is a unitary $\Z_2$ operator as the weak symmetry $G=\Z_2$. 
Such FDLUC $\mathcal{U}$ can be simulated with a continuous unitary
evolution with local Hamiltonian $H_I(t)$ for a finite $t\in[0,1]$ time \cite{PhysRevB.82.155138}. This local  Hamiltonian $H_I(t)$ preserves $D_{4n}$ symmetry for the whole evolution, i.e.
\[
\begin{split}
[U\otimes I_E,H_I(t) ]=[\eta^2\otimes I_E,H_I(t) ]=[\eta\otimes \eta_{E},H_I(t)]=0.\nonumber
\end{split}
\]
After gauging $\Z^U_2$ symmetry, this $\mathcal{U}$ maps to a dual FDLUC $\mathcal{U}'$ which can be simulated using a dual local  Hamiltonian $H'_I(t)$:
\[
(\mathcal{D}\otimes I_E) H_I(t)=H'_I(t)(\mathcal{D}\otimes I_E). 
\]
This equation reflects the fact that the system carries a strong symmetry charge while the environment is neutral under strong symmetry. 

This dual  Hamiltonian satisfies
\[
\begin{split}
&[U_{e/o}\otimes I_E,H'_I(t)]=[\mathsf{W}_{2k}\otimes I_E,H'_I(t)]\\=&[\mathsf{W}_{2k-1}\otimes \eta_E, H'_I(t)]=0.\nonumber
\end{split}
\]
This implies that $\mathcal{U}'$ is locally symmetric under $\text{Rep}(D_{4n})$, as the $U_{e/o}\otimes I_E$, $\mathsf{W}_{2k}\otimes I_E$ and $\mathsf{W}_{2k-1}\otimes \eta_E$ form the fusion category Rep($D_{4n}$). Therefore this $\mathcal{U}'$ generate a $\text{Rep}(D_{4n})$ locally-symmetric FDLC $\mathcal{N'}$:
\[
\begin{split}
\mathcal{D}\mathcal{N}[\rho]&=\Tr_E\{(\mathcal{D}\otimes I_E)\mathcal{U}\rho\otimes|0\rangle_E\langle 0|_E\mathcal{U}^\dagger\}\\&=\Tr_E\{\mathcal{U}'\rho'\otimes|0\rangle_E\langle 0|_E\mathcal{U}'^\dagger\}\mathcal{D}=\mathcal{N'}[\rho']\mathcal{D},
\end{split}
\]
where $\mathcal{D}\rho=\rho'\mathcal{D}$ defines the dual mixed state. This implies that each $D_{4n}$ local-symmetric FDLC is mapped to, under gauging $\Z^U_2$, a $\text{Rep}(D_{4n})$ locally-symmetric FDLC. Moreover, by applying the ungauging operation $\mathcal{D}^{\dagger}$ (equivalent to gauging the quantum symmetry of dual systems), we can obtain the converse version of the above statement. As a result, two states lie in the same $D_{4n}$ mixed-state phase---i.e., they are two-way channel connected by locally symmetric channels---if and only if their dual states obtained by gauging $\mathbb{Z}_2^{U}$ lie in the same $\mathrm{Rep}(D_{4n})$ mixed-state phase defined by two-way channel connectivity as well.

Here we remark that the above argument can be straightforwardly generalized to any duality, by gauging strong symmetries, between models with generalized symmetries $\mathcal C_1$ and $\mathcal C_2$. One crucial point is that the gauging duality acts only on the system while leaving the environment invariant. Moreover, the duality connects the symmetry actions on the total systems  $S\cup E$ in the purification on both sides, where these symmetry actions also form representations of  $\mathcal C_1$ and $\mathcal C_2$, respectively. Combining these two observations, one can obtain that the gauging duality induces a one-to-one correspondence between $\mathcal C_1$ locally symmetric FDLCs and $\mathcal C_2$ locally symmetric FDLCs on two sides.

\subsection{Classification of Rep($D_{4n}$) ANISPT}
In this section, we further study the classification of Rep($D_{4n}$) ANISPT in $(1+1)d$, which corresponds to the second case in Sec.~\ref{sec:repd2mopen} with the weak symmetry $\mathsf{W}_{2k-1}$ and strong symmetry $\mathsf{W}_{2k},U_{e/o}$, whose symmetry action is  \eqref{eq: symmetry conditions-1}. We summarize our results in Tab.~\ref{tab:SPTclassification}, where we only list the number of each class because NISPTs in general do not admit a group structure~\cite{Seifnashri:2024dsd,Cao:2025qhg}.

As proved in the last section, the mixed-state phases with dual Rep($D_{4n}$) symmetry can be studied by the corresponding mixed-state phases with original $D_{4n}$ symmetry. In particular, Rep($D_{4n}$) ANISPTs correspond to SSB phases in the $D_{4n}$ symmetric systems  where the $\mathbb Z_2^U$ is strongly broken. The reason is as follows: If we only focus on the $\mathbb Z_2^U$ sub-symmetry and its dual $\hat{\mathbb Z}_2$ sub-symmetry, the KW-duality establishes the bijection between the only $\mathbb Z_2^U$ strong SSB phase and the $\hat{\mathbb Z}_2$ ASPT phase, i.e., the trivial phase. Considering other symmetries, we make a finer classification on both sides, and the broken symmetry feature will not change.

To classify $\mathbb Z_2^U$-SSB phases in the $D_{4n}$ symmetric system, we need to identify the unbroken symmetry group $H$ and the possible ASPT phases protected by $H$. In this case, there are two choices of unbroken symmetry. One is $H=\mathbb{Z}_{2n}^{\eta}$, the other is $H=\mathbb Z_{n}^{\eta^2}\rtimes\mathbb Z_2^{\eta U}=\{\eta^2|\eta^{2n}=1\}\rtimes \{1,\eta U\}$. 

    

\begin{table}[!tbp]
    \centering
    \small 
    \renewcommand{\arraystretch}{1.08}
    \setlength{\tabcolsep}{3.5pt} 

    \begin{tabular}{|c|c|c|c|c|}
        \hline
        \multicolumn{2}{|c|}{$\mathrm{Rep}(D_{4n})$ symmetry}
        & \multicolumn{2}{c|}{ANISPT}
        & IANISPT \\
        \hline
        \multicolumn{2}{|c|}{Unbroken group $H \subset D_{4n}$}
        & $\mathbb{Z}_{2n}^{\eta}$
        & $\mathbb{Z}_{n}^{\eta^{2}}\rtimes \mathbb{Z}_{2}^{\eta U}$
        & $\mathbb{Z}_{2n}^{\eta}$ \\
        \hline
        \multirow{2}{*}{\makecell[c]{Number\\of phases}}
        & $n$ odd  & 1 & 1 & 0 \\
        \cline{2-5}
        & $n$ even & 1 & 2 & 1 \\
        \hline
    \end{tabular}

    \caption{Classification of Rep$(D_{4n})$ ANISPTs. Each branch in the second row corresponds to the unbroken group of $D_{4n}$ symmetric systems before gauging and we show only the number of ANISPT in each branch.}
    \label{tab:SPTclassification}
\end{table}

Next, let us further consider the attached $H$-ASPT phases.  Recall that, before gauging $\Z^U_2$, the strong symmetry subgroup of $D_{4n}$ is $K=\mathbb Z_{n}^{\eta^2}\rtimes\mathbb Z_2^{U}=\{\eta^2|\eta^{2n}=1\}\rtimes \{1,U\}$ and the weak symmetry is $G=\mathbb Z_2$.

\begin{enumerate}
    \item For $H=\mathbb{Z}_{2n}^{\eta}$, the strong subgroup is $H\cap K=\mathbb{Z}_{n}^{\eta^2}$ and the weak symmetry group is $\mathbb{Z}_2$. For $n$ is odd, there is only a trivial ASPT phase with $H$ symmetry~\cite{PhysRevX.13.031016}. Correspondingly, after gauging, we get an ANISPT phase with Rep($D_{4n}$) symmetry. We find the pure state correspondence for this phase can be represented as a symmetric product state on the lattice~\cite{Cao:2025qhg}, thus we denote it as the trivial phase.

    For $n$ is even, the classification of $H$-ASPT includes a trivial phase and an intrinsic ASPT (more details in section \ref{sec: class of dASPT}). This $H$-IASPT phase has no pure-state counterpart. Correspondingly, gauging the $\Z^U_2$-SSB phase with the $H$-IASPT order gives rise to a Rep($D_{4n}$) intrinsic ANISPT (IANISPT). 

    \item When $H=\mathbb Z_{n}^{\eta^2}\rtimes\mathbb Z_2^{\eta U}=\{\eta^2|\eta^{2n}=1\}\rtimes \{1,\eta U\}$ where $\mathbb Z_{n}^{\eta^2}$ is strong and  $\Z_2^{\eta U}$ is weak, there are one non-intrinsic $H$-ASPT when $n$ is odd and two non-intrinsic $H$-ASPT when $n$ is even, which both also have pure-state counterparts \footnote{The pure state SPT with $H$ symmetry is jointly protected by $\Z_{n}^{\eta^2}$ and $\Z_2^{\eta U}$, characterized by the nontrivial commutation relation between $\eta^2$ and $U$ on the edge modes. Thus when $\Z^U_2$ is weak, the classification of the ASPT phase remains the same, as it is still determined by this commutation relation on the edge modes.}. Moreover, we also prove that there is no $H$-IASPT phase in appendix \ref{app1}. After gauging back in the Rep($D_{4n}$) symmetric system, the $\Z^U_2$-SSB phases with these $H$-ASPT order  will be mapped to Rep($D_{4n}$) non-intrinsic ANISPTs. When $n$ is even, since the pure state counterpart is denoted as even and odd SPTs \cite{Cao:2025qhg}, we also denote these two ANISPT as even and odd ANISPT phases. 
    
\end{enumerate}

In summary, for Rep($D_{4n}$) symmetry,  there is an intrinsic  ANISPT, together with 3 non-intrinsic  ANISPTs when $n$ is even, and two  non-intrinsic  ANISPTs when $n$ is odd.

\section{Lattice models of non-intrinsic Rep$(D_{8})$ ANISPT}
In this section, we construct lattice models realizing three non-intrinsic Rep$(D_{8})$ ANISPTs. As discussed in the previous section, these ANISPTs correspond to distinct $\Z^U_2$-SSB phases in the $D_8$ symmetric systems before gauging. In particular, there is one $H$-ASPT for $H=\Z_4$ and there are two $H$-ASPT for $H=\Z_2\times \Z_2$. We will first construct $D_8$ symmetric Hamiltonian realizing $\Z^U_2$-SSB phases and then apply $D_8$ symmetric channels to the $\Z^U_2$-even ground states. For each case, we give a detailed analysis of the density matrices and phase diagrams using an effective approach. Next, we apply the KW duality to these $\Z^U_2$-SSB models, which amounts to gauging the $\Z^U_2$ symmetry. The resulting dual Hamiltonian and channels realize Rep$(D_{8})$ ANISPTs. We then give a detailed analysis of density matrices and phase diagrams. The generalization to general non-intrinsic Rep$(D_{4n})$ ANISPTs is straightforward and we leave it in App.~\ref{app:anispt}.

\subsection{Trivial Rep$(D_{8})$ ANISPT phase}\label{sec: trivial}
Let us consider the case where $H=\Z^{\eta}_4$ with $H\cap K=\mathbb{Z}_{2}^{\eta^2}$. We assume the length $L\in 4\Z$ and group 4 spins into a unit. Then we consider the following model before gauging. The $D_8$ symmetry is given by $\eqref{eq: symmetry action}$ for $n=2$, generated by $90\degree$ spin rotation $\eta$ and spin flip $U$. The symmetric Hamiltonian is
\begin{align}\label{eq: SSB Hal}
    \begin{split}
      &H=-\sum_{j=1}^L\sigma^z_{j}\sigma^z_{j+1}\\
    &-\frac{h}{2}\sum_{i=1}^{L/4} (\sigma^x_{4i-3}\sigma^x_{4i-2}\sigma^x_{4i-1}\sigma^x_{4i}+\sigma^y_{4i-3}\sigma^y_{4i-2}\sigma^y_{4i-1}\sigma^y_{4i}) , 
    \end{split}
\end{align}
and we apply the channel $\mathcal{N}=\otimes^{L/4}_{i=1}\mathcal{N}^{x}_{i}\circ \mathcal{N}^{y}_{i}$ to the $\Z^U_2$-even ground state $\rho_0=|\text{G.S.}\rangle_{e}\langle\text{G.S.}|$, where
\begin{equation}\label{eq:quantumchannel}
\begin{split}
    &\mathcal{N}^{x}_{i}[\rho]=(1-p)\rho+p\sigma^x_{4i-3}\sigma^x_{4i-2}\sigma^x_{4i-1}\sigma^x_{4i}\rho \sigma^x_{4i-3}\sigma^x_{4i-2}\sigma^x_{4i-1}\sigma^x_{4i},\\
    &\mathcal{N}^{y}_{i}[\rho]=(1-p)\rho+p\sigma^y_{4i-3}\sigma^y_{4i-2}\sigma^y_{4i-1}\sigma^y_{4i}\rho \sigma^y_{4i-3}\sigma^y_{4i-2}\sigma^y_{4i-1}\sigma^y_{4i}. 
\end{split}
\end{equation}
It is easy to check that the channel is strongly invariant under $\eta^2$ and weakly invariant under $\eta$.

When $h=0$ and $p=0$, the system is described by the GHZ state:
\begin{equation}
    |\text{GHZ}\rangle\propto |\uparrow\uparrow\uparrow\cdots \uparrow\rangle+|\downarrow\downarrow\downarrow\cdots \downarrow\rangle,
\end{equation}
which belongs to the $\Z^U_2$-SSB phase.

To solve the resulting density matrix after adding this channel, namely $\mathcal{N}[\rho_0]$, we notice that the local operators $\sigma^z_{4i-3}\sigma^z_{4i-2}$, $\sigma^z_{4i-2}\sigma^z_{4i-1}$, $\sigma^z_{4i-1}\sigma^z_{4i}$ commute with the Hamiltonian and channel. Since the GHZ state is an eigenstate of these local operators with eigenvalue 1, the resulting density matrix $\mathcal{N}[\rho_0]$ also has eigenvalue 1. Thus $\mathcal{N}[\rho_0]$ must be expanded in the subspace with this constraint $\sigma^z_{4i-3}\sigma^z_{4i-2}=\sigma^z_{4i-2}\sigma^z_{4i-1}=\sigma^z_{4i-1}\sigma^z_{4i}=1$. In each unit cell, this subspace is expanded by 
    \begin{align}
        \begin{split}
         |\tilde{\uparrow}\rangle_i&=|\uparrow\uparrow\uparrow\uparrow\rangle_{4i-3,4i-2,4i-1,4i},\\ |\tilde{\downarrow}\rangle_i&=|\downarrow\downarrow\downarrow\downarrow\rangle_{4i-3,4i-2.4i-1,4i}.
        \end{split}
    \end{align}
    In this subspace, the effective Hamiltonian is the transverse field lsing model
\begin{equation}
H_{\text{eff}}=-\sum_{i=1}^{L/4} \tilde{\sigma}^z_i \tilde{\sigma}^z_{i+1}-h\sum_{i=1}^{L/4} \tilde{\sigma}^x_i,
    \end{equation}
    and the above channel acts effectively as
    \begin{equation}
    \mathcal{N}^{x/y}_{i}[\rho]\to \tilde{\mathcal{N}}_{i}[\tilde{\rho}]=(1-p)\tilde{\rho}+p\tilde{\sigma}^x_{i}\tilde{\rho} \tilde{\sigma}^x_i.
    \end{equation} 
The $\Z^U_2$ symmetry is effectively $U= \prod \tilde{\sigma}^x_i$. We denote the resulting effective mixed state and effective Pauli matrix in the subspace as $\tilde{\rho}$ and $\tilde{\sigma}^{x/y/z} $.

\begin{figure}
    \centering
    \resizebox{0.95\linewidth}{!}{
    \begin{tikzpicture}[x=0.8cm, y=0.8cm]
    \begin{scope}[shift={(0, 0)}]
    \draw[thick, ->] (0,0) -- (12,0);
    \draw[thick, ->] (0,0) -- (0,6);
    \fill (6,0) circle (3pt);
	   \fill (0,0) circle (3pt);
	    \node[below] at (12,0) {$h=\infty$};
	    \node[below] at (6,0) {$h=1$};
	    \node[below] at (0,0) {$0$};
	    \node[left] at (0,6) {$p=\frac{1}{2}$};
	    \draw[very thick, dashed] (6,0) -- (6,6);
	    \draw[very thick, dashed] (0,6) -- (12,6);
	    \node[above] at (3,6) {$\Z^U_2$ SWSSB};
	  
	    \node at (3,3) {\begin{tabular}{c}
	         $\Z^U_2$ SSB $(H=\Z^{\eta}_4)$
	    \end{tabular} };
	     \node at (9, 3) {\begin{tabular}{c}
	         Trivial
	    \end{tabular} };
	   
	    \node[above] at (6,6) {Transition};
	    \node[above] at (9,6) {Trivial};
	    \end{scope}
	   \begin{scope}[shift={(0, -8.5)}]
        \draw[thick, ->] (0,0) -- (12,0);
    \draw[thick, ->] (0,0) -- (0,6);
    \fill (6,0) circle (3pt);
	   \fill (0,0) circle (3pt);
	    \node[below] at (12,0) {$h=\infty$};
	    \node[below] at (6,0) {$h=1$};
	    \node[below] at (0,0) {$0$};
	    \node[left] at (0,6) {$p=\frac{1}{2}$};
	    \draw[very thick, dashed] (6,0) -- (6,6);
	    \draw[very thick, dashed] (0,6) -- (12,6);
	    \node[above] at (3,6) {$\Z^U_2$ SWSSB};
	  
	    \node at (3,3) {\begin{tabular}{c}
	         Trivial ANISPT
	    \end{tabular} };
	     \node at (9, 3) {\begin{tabular}{c}
	        $\Z^U_2$ SSB
	    \end{tabular} };
	   
	    \node[above] at (6,6) {Transition};
	    \node[above] at (9,6) {$\Z^U_2$ SSB};
    \end{scope}
    \end{tikzpicture}
    }
    \caption{The Phase diagram of model with Hamiltonian \eqref{eq: SSB Hal} and channel \eqref{eq:quantumchannel}  before gauging and the phase diagram of the model with Hamiltonian \eqref{eq: dual Hal} and channel \eqref{eq: dual channel} after gauging. }
    \label{fig:SSBpert-1}
\end{figure}
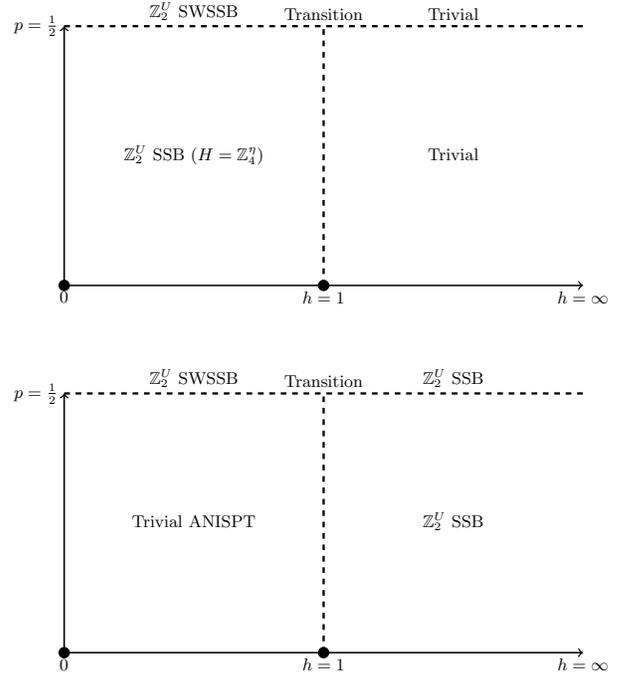

The phase diagram of this effective model has been studied in Ref.~\cite{luo2025topological} and thus we can obtain the phase diagram accordingly in the first figure of Fig.~\ref{fig:SSBpert-1}.
\begin{enumerate}
    \item When $0\leq h <1$ and $0\le p<1/2$, the system belongs to the $\Z^U_2$ SSB phase. 
    \item When $1<h$ and $0\le p\le 1/2$, the system belongs to the $D_8$ symmetric trivial phase.
    \item When $0\leq h <1$ and $p=1/2$, the system belongs to the $\Z^U_2$ strong-to-weak SSB (SWSSB) phase.
\end{enumerate}
Let us focus on two interesting lines where $h=0$ and $p=1/2$. 

When $h=0$, the effective density matrix is given by 
$\tilde{\rho}=\tilde{\mathcal{N}}[|\text{GHZ}\rangle\langle\text{GHZ}| ]$
and we can obtain
\begin{equation}
    \text{Tr}(\tilde{\rho} \tilde{\sigma}^z_i\tilde{\sigma}^z_j)=(1-2p)^2.
\end{equation}
Since $\tilde{\sigma}^z_j$ is charged under $\Z^U_2$ symmetry, the density matrix realizes the strong $\Z^U_2$ SSB phase when $p\ne \frac{1}{2}$. Indeed, when $p=\frac{1}{2}$, the original mixed state $\rho=\mathcal{N}[\rho_0]$ is given by
\begin{equation}\label{eq:SWSSB-1}
    \frac{\mathbb{I}+\prod {\sigma}^x_i}{2}\prod_i \frac{\mathbb{I}+\sigma^z_{4i-3}\sigma^z_{4i-2}}{2}\frac{\mathbb{I}+\sigma^z_{4i-2}\sigma^z_{4i-1}}{2}\frac{\mathbb{I}+\sigma^z_{4i-1}\sigma^z_{4i}}{2},
\end{equation}
   which realizes the $\Z^U_2$ SWSSB phase. The projection to $\Z^U_2$ even-sector in the first part of the above density matrix characterizes  the $\Z^U_2$ SWSSB feature.

Let us consider another line with $p=\frac{1}{2}$, where the model can be characterized by the disorder parameter 
\[\label{eq: disorder-1}
\text{Tr}(\rho \prod^j_{k=i}\sigma^x_{4k-3}\sigma^x_{4k-2}\sigma^x_{4k-1}\sigma^x_{4k})=\text{Tr}(\tilde{\rho} \prod^j_{k=i}\tilde{\sigma}^x_{k}) 
\]
and 
the Wightman correlation functions:
\[\label{eq: Wightman-1}
\text{Tr}(\sqrt{\rho} \sigma^z_{4i}\sigma^z_{4j}\sqrt{\rho} \sigma^z_{4i}\sigma^z_{4j})=\text{Tr}(\sqrt{\tilde{\rho}} \tilde{\sigma}^z_{i}\tilde{\sigma}^z_{j}\sqrt{\tilde{\rho}} \tilde{\sigma}^z_{i}\tilde{\sigma}^z_{j}).
\]
According to \cite{luo2025topological}, when $0 \leq h<1$, the disorder parameter exhibits exponential decay and the Wightman correlation function is long range in  $\Z^U_2$ SWSSB phase. When $h>1$, the disorder parameter is long range and the Wightman correlation function exhibits exponential decay in  the trivial phase. At mixed-state phase transition $h=1$, both two parameter have a power-law behavior $\frac{1}{|i-j|^{\frac{1}{4}}}$.

To obtain the ANISPT phase, we perform the KW duality on this model. The dual Hamiltonian is
\[\label{eq: dual Hal}
H=-\sum_{i}\sigma^x_{i}-\frac{h}{2}\sum_i \sigma^z_{4i-4}\sigma^z_{4i}(1+\sigma^x_{4i-3}\sigma^x_{4i-1})
\]
and the dual channel is
\begin{equation}\label{eq: dual channel}
\begin{split}
    &\mathcal{N}^{x}_{i}[\rho]=(1-p)\rho+p\sigma^z_{4i-4}\sigma^z_{4i}\rho \sigma^z_{4i-4}\sigma^z_{4i},\\
    &\mathcal{N}^{y}_{i}[\rho]=(1-p)\rho+p\sigma^z_{4i-4}\sigma^z_{4i}\sigma^x_{4i-3}\sigma^x_{4i-1}\rho \sigma^z_{4i-4}\sigma^z_{4i}\sigma^x_{4i-3}\sigma^x_{4i-1}.\nonumber
\end{split}
\end{equation}

The phase diagram after KW duality is shown in the second picture in Fig.~\ref{fig:SSBpert-1}.
\begin{enumerate}
    \item When $0\leq h <1$ and $0\le p<1/2$, the system belongs to the trivial ASPT phase. This is because when $h=p=0$ the mixed-state is the product state: $|+++\cdots+\rangle\langle+++\cdots+|$, which is the reason for the denotation `trivial ANISPT' phase. 
    \item When $1<h$ and $0\le p\le 1/2$, the system belongs to the $\Z^U_2$ SSB phase where $\Z^{U}_2$ is the diagonal subgroup generated by $U_e U_o=\prod_j \sigma^x_j$.
    \item When $0\leq h <1$ and $p=1/2$, the system belongs to $\Z^U_2$ SWSSB phase. In particular, the mixed state at $p=\frac{1}{2}$ and $h=0$ is obtained by applying KW duality on the mixed-state \eqref{eq:SWSSB-1} and is given by
\begin{equation}
     \rho=\frac{\mathbb{I}+\prod {\sigma}^x_i}{2}\prod_i \frac{\mathbb{I}+\sigma^x_{4i-3}}{2}\frac{\mathbb{I}+\sigma^z_{4i-2}}{2}\frac{\mathbb{I}+\sigma^x_{4i-1}}{2}.
\end{equation}
The projection in the first part of the above density matrix characterizes  the $\Z^U_2$ SWSSB feature.
\end{enumerate}
Moreover, in the line with $p=\frac{1}{2}$, the mixed state after gauging can be characterized by the dual order parameter 
\[
\text{Tr}(\rho \sigma^z_{4i}\sigma^z_{4j-1}) 
\]
and 
the dual Wightman disorder parameter:
\[
\text{Tr}(\sqrt{\rho} \mu_{i,j}\sqrt{\rho} \mu_{i,j}),
\]
where $\mu_{i,j}=\prod^j_{k=i}\sigma^x_{4k}\sigma^x_{4k+1}\sigma^x_{4k+2}\sigma^x_{4k+3}$. These two quantities follow from applying the KW duality on \eqref{eq: disorder-1} and \eqref{eq: Wightman-1}.
When $0\leq h<1$, the dual order parameter exhibits exponential decay and the dual Wightman disorder parameter is long range in  $\Z^U_2$ SWSSB phase. When $h>1$, the dual order parameter is long range and the dual Wightman correlation function exhibits exponential decay in $\Z^U_2$-SSB.  At mixed-state phase transition $h=1$, both two parameter have a power-law behavior $\frac{1}{|i-j|^{\frac{1}{4}}}$.

	  
	   

\subsection{Even and odd Rep$(D_{8})$ ANISPT phases}\label{sec:evenodd}
Let us continue to construct even and odd Rep($D_8$) ANISPT phases.  By KW duality, these two phases correspond to two $\Z^U_2$-SSB phases where unbroken group $H=\Z^{\eta^2}_2\times \Z^{\eta U}_2$ and $\Z^{\eta^2}_2$ is strong and $\Z^{\eta U}_2$ is weak.  
In this case, the unbroken group $H$ allows two distinct ASPT phases and we will show they are related by one-site translation. 
\begin{figure}
    \centering
    \resizebox{0.95\linewidth}{!}{
    \begin{tikzpicture}[x=0.8cm, y=0.8cm]
    \begin{scope}[shift={(0, 0)}]
    \draw[thick, ->] (0,0) -- (12,0);
    \draw[thick, ->] (0,0) -- (0,6);
    \fill (6,0) circle (3pt);
	   \fill (0,0) circle (3pt);
	    \node[below] at (12,0) {$h=\infty$};
	    \node[below] at (6,0) {$h=1$};
	    \node[below] at (0,0) {$0$};
	    \node[left] at (0,6) {$p=\frac{1}{2}$};
	    \draw[very thick, dashed] (6,0) -- (6,6);
	    \draw[very thick, dashed] (0,6) -- (12,6);
	    \node[above] at (3,6) {$\Z^U_2$ SWSSB};
	  
	    \node at (3,3) {\begin{tabular}{c}
	         $\Z^U_2$ SSB $(H=\mathbb Z_{2}^{\eta^2}\times\mathbb Z_2^{\eta U})$
	    \end{tabular} };
	     \node at (9, 3) {\begin{tabular}{c}
	         Trivial
	    \end{tabular} };
	   
	    \node[above] at (6,6) {Transition};
	    \node[above] at (9,6) {Trivial};
	    \end{scope}
	   \begin{scope}[shift={(0, -8.5)}]
        \draw[thick, ->] (0,0) -- (12,0);
    \draw[thick, ->] (0,0) -- (0,6);
    \fill (6,0) circle (3pt);
	   \fill (0,0) circle (3pt);
	    \node[below] at (12,0) {$h=\infty$};
	    \node[below] at (6,0) {$h=1$};
	    \node[below] at (0,0) {$0$};
	    \node[left] at (0,6) {$p=\frac{1}{2}$};
	    \draw[very thick, dashed] (6,0) -- (6,6);
	    \draw[very thick, dashed] (0,6) -- (12,6);
	    \node[above] at (3,6) {$\Z^U_2$ SWSSB};
	  
	    \node at (3,3) {\begin{tabular}{c}
	         Even ANISPT
	    \end{tabular} };
	     \node at (9, 3) {\begin{tabular}{c}
	         $\Z^U_2$-SSB
	    \end{tabular} };
	   
	    \node[above] at (6,6) {Transition};
	    \node[above] at (9,6) {$\Z^U_2$-SSB};
    \end{scope}
    \end{tikzpicture}
    }
    \caption{The Phase diagram of model with Hamiltonian \eqref{eq: SSB HAL2} and channel \eqref{eq:quantumchannel}  before gauging and the phase diagram of the model with Hamiltonian \eqref{eq: even HAL} and channel \eqref{eq: dual channel} after gauging.}
    \label{fig:SSBpert-2}
\end{figure}
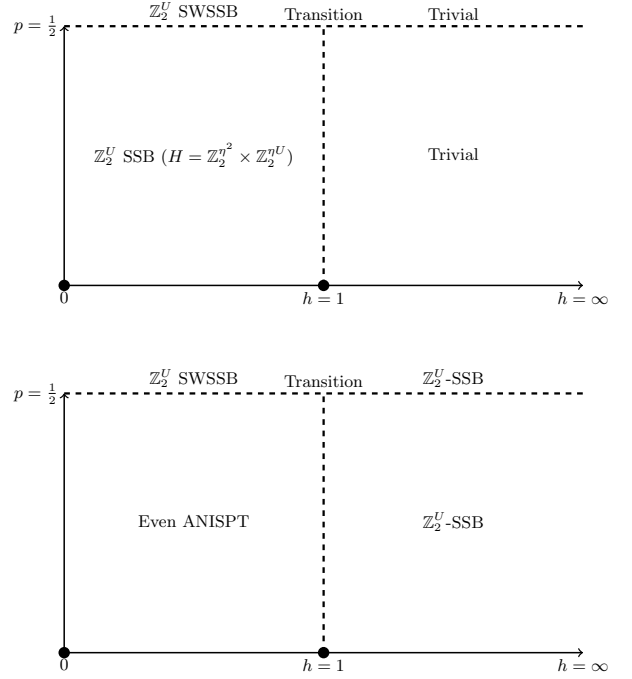
Let us consider the following $D_8$ symmetric model which is dual to the even ANISPT phase. The Hamiltonian is
\[\label{eq: SSB HAL2}
\begin{split}
H=&-\sum_{j=1}^{L/2} \sigma^z_{2j-1}\sigma^z_{2j}\\
&+\frac{1}{2}\sum_{j=1}^{L/2}(\sigma^y_{2j-1}\sigma^x_{2j}\sigma^x_{2j+1}\sigma^y_{2j+2}+\sigma^x_{2j-1}\sigma^y_{2j}\sigma^y_{2j+1}\sigma^x_{2j+2})\\&-\frac{h}{2}\sum_{i=1}^{L/4} (\sigma^x_{4i-3}\sigma^x_{4i-2}\sigma^x_{4i-1}\sigma^x_{4i}+\sigma^y_{4i-3}\sigma^y_{4i-2}\sigma^y_{4i-1}\sigma^y_{4i}),\nonumber
\end{split}
\]
and we apply the same channel \eqref{eq:quantumchannel} as the previous case to the $\Z^U_2$-even ground state $\rho_0=|\text{G.S.}\rangle_{e}\langle\text{G.S.}|$.
To obtain the mixed state $\mathcal{N}[\rho_0]$, similarly, we notice that local terms $\sigma^z_{2i-1}\sigma^z_{2i}$ and $\sigma^y_{4i-3}\sigma^x_{4i-2}\sigma^x_{4i-1}\sigma^y_{4i}$ commute with Hamiltonian and channel. In particular, when $h=p=0$, the ground state should be an eigenstate of $\sigma^z_{2i-1}\sigma^z_{2i}$ and $\sigma^y_{4i-3}\sigma^x_{4i-2}\sigma^x_{4i-1}\sigma^y_{4i}$  with eigenvalue $1$ and $-1$, since these two local terms correspond to the first and second term of Hamiltonian. Thus $\mathcal{N}[\rho_0]$ must be expanded in the subspace with the constraint $\sigma^z_{2i-1}\sigma^z_{2i}=-\sigma^y_{4i-3}\sigma^x_{4i-2}\sigma^x_{4i-1}\sigma^y_{4i}=1$. As a result, we denote 
\begin{equation}
\begin{split}
&|\tilde{\uparrow}\rangle_i=\frac{1}{\sqrt{2}}(|\uparrow\uparrow\uparrow\uparrow\rangle_{4i-3,4i-2,4i-1.4i}+|\downarrow\downarrow\downarrow\downarrow\rangle_{4i-3,4i-2.4i-1,4i}), \\
&|\tilde{\downarrow}\rangle_i=\frac{1}{\sqrt{2}}(|\uparrow\uparrow\downarrow\downarrow\rangle_{4i-3,4i-2.4i-1,4i}-|\downarrow\downarrow\uparrow\uparrow\rangle_{4i-3,4i-2.4i-1,4i}),\nonumber
\end{split}
    \end{equation}
which span the subspace in each unit cell.
    
In this subspace, the effective Hamiltonian is the transverse field Ising model
\begin{equation}
H_{\text{eff}}=-\sum_{i=1}^{L/4} \tilde{\sigma}^y_i \tilde{\sigma}^y_{i+1}-h\sum_{i=1}^{L/4} \tilde{\sigma}^z_i
    \end{equation}and the above channel acts effectively as
    \begin{equation}
    \mathcal{N}^{x/y}_{i}[\rho]\to \tilde{\mathcal{N}}_i[\tilde{\rho}]=(1-p)\tilde{\rho}+p\tilde{\sigma}^z_{i}\tilde{\rho} \tilde{\sigma}^z_i.
    \end{equation}    
Effectively, $\Z^U_2$ symmetry $U$ becomes $\prod \tilde{\sigma}^z_i$, and $\Z^{\eta}_4$ symmetry $\eta$ becomes $\prod \tilde{\sigma}^z_i$. Thus $\eta^2$ and $U \eta $ are effectively the identity in this subspace and are unbroken. The phase diagram of this model is shown in the first figure of Fig.~\ref{fig:SSBpert-2}, which is similar to the first figure of Fig.~\ref{fig:SSBpert-1}. The only difference is that the unbroken group of $\Z^U_2$ SSB phase is $H=\Z^{\eta^2}_2\times\Z^{\eta U}_2$. We also consider two interesting lines with $h=0$ and $p=1/2$.

 When $h=0$, the effective density matrix is given by 
$\tilde{\rho}=\tilde{\mathcal{N}}[|\text{GHZ}_y\rangle\langle\text{GHZ}_y| ]$, where
\[
|\text{GHZ}_y\rangle\propto\otimes^L_{i=1}|\tilde{\sigma}^y_i=1\rangle+\otimes^L_{i=1}|\tilde{\sigma}^y_i=-1\rangle,
\]
and we can obtain
\begin{equation}
    \text{Tr}(\tilde{\rho} \tilde{\sigma}^y_i\tilde{\sigma}^y_j)=(1-2p)^2.
\end{equation}
Thus the density matrix is in the strong $\Z^U_2$-SSB phase when $p\ne \frac{1}{2}$. Moreover, in the region of $\Z^U_2$-SSB phase when $0\leq h<1$ and $p<1/2$, we have the nonzero correlation function
\begin{align}\label{eq:corr-2}
    \begin{split}
      &\text{Tr}(\rho (\sigma^x_{4i-1}\sigma^y_{4i}+\sigma^y_{4i-1}\sigma^x_{4i})(\sigma^y_{4j-1}\sigma^x_{4j}+\sigma^x_{4j-1}\sigma^y_{4j}))\\
      &=4\text{Tr}(\tilde{\rho} \tilde{\sigma}^y_i\tilde{\sigma}^y_j)\ne 0.  
    \end{split}
\end{align}
 The above local operator is odd under $\Z^U_2$ and $\Z^{\eta}_4$ symmetry.

When $p=\frac{1}{2}$ and $h=0$, the original mixed state $\rho=\mathcal{N}[\rho_0]$ is given by
    \begin{equation}
    \frac{\mathbb{I}+\prod {\sigma}^x_i}{2}\prod_i \frac{\mathbb{I}+\sigma^z_{4i-3}\sigma^z_{4i-2}}{2}\frac{\mathbb{I}-\sigma^y_{4i-3}\sigma^x_{4i-2}\sigma^x_{4i-1}\sigma^y_{4i}}{2}\frac{\mathbb{I}+\sigma^z_{4i-1}\sigma^z_{4i}}{2}
\end{equation}
   which belongs to the $\Z^U_2$ SWSSB phase.

When $p=\frac{1}{2}$, the model can be characterized by the disorder parameter 
\[\label{eq: disorder-2}
\text{Tr}(\rho \prod^j_{k=i}\sigma^x_{4k-3}\sigma^x_{4k-2}\sigma^x_{4k-1}\sigma^x_{4k})=\text{Tr}(\tilde{\rho} \prod^j_{k=i}\tilde{\sigma}^x_{k}), 
\]
and 
the following Wightman correlation functions:
\begin{widetext}
\[\label{eq: weightman-2}
\text{Tr}(\sqrt{\rho} (\sigma^x_{4i-1}\sigma^y_{4i}+\sigma^y_{4i-1}\sigma^x_{4i})(\sigma^y_{4j-1}\sigma^x_{4j}+\sigma^x_{4j-1}\sigma^y_{4j})\sqrt{\rho} (\sigma^x_{4i-1}\sigma^y_{4i}+\sigma^y_{4i-1}\sigma^x_{4i})(\sigma^y_{4j-1}\sigma^x_{4j}+\sigma^x_{4j-1}\sigma^y_{4j}))=\text{Tr}(\sqrt{\tilde{\rho}} \tilde{\sigma}^y_{i}\tilde{\sigma}^y_{j}\sqrt{\tilde{\rho}} \tilde{\sigma}^y_{i}\tilde{\sigma}^y_{j}).
\]
\end{widetext}
According to \cite{luo2025topological}, when $0<h<1$, the disorder parameter exhibits exponential decay and the Wightman correlation function is long range in  $\Z^U_2$ SWSSB phase. When $h>1$, the disorder parameter is long range and the Wightman correlation function exhibits exponential decay in the trivial phase. At mixed-state phase transition $h=1$, both two parameter have a power-law behavior $\frac{1}{|i-j|^{\frac{1}{4}}}$.

After gauging $\Z^U_2$ symmetry, the Hamiltonian  becomes
\[\label{eq: even HAL}
\begin{split}
H=&-\sum_{j=1}^{L/2} \sigma^x_{2j-1}-\frac{1}{2}\sum_{j=1}^{L/2}\sigma^z_{2j-2}\sigma^x_{2j}\sigma^z_{2j+2}(\sigma^x_{2j-1}\sigma^x_{2j+1}+1)\\&-\frac{h}{2}\sum_{i=1}^{L/4} \sigma^z_{4i-4}\sigma^z_{4i}(1+\sigma^x_{4i-3}\sigma^x_{4i-1}),
\end{split}
\]
with the following channel:
\begin{equation}\label{eq: even channel}
\begin{split}
    &\mathcal{N}^{x}_{i}[\rho]=(1-p)\rho+p\sigma^z_{4i-4}\sigma^z_{4i}\rho \sigma^z_{4i-4}\sigma^z_{4i},\\
    &\mathcal{N}^{y}_{i}[\rho]=(1-p)\rho+p\sigma^z_{4i-4}\sigma^z_{4i}\sigma^x_{4i-3}\sigma^x_{4i-1}\rho \sigma^z_{4i-4}\sigma^z_{4i}\sigma^x_{4i-3}\sigma^x_{4i-1}.
    \end{split}
    \end{equation}
Similarly, there are local terms $\sigma^x_{2i-1}$ and $\sigma^z_{2i-2}\sigma^x_{2i}\sigma^z_{2i+2}$ commuting with the Hamiltonian and channel. Thus the mixed state $\mathcal{N}[\rho'_0]$, where $\rho'_0=|\text{G.S.}\rangle_{e}\langle\text{G.S.}|$ is the ground state of \eqref{eq: even HAL},  must be expanded in the subspace with the constraint $\sigma^x_{2i-1}=\sigma^z_{2i-2}\sigma^x_{2i}\sigma^z_{2i+2}=1$. This implies $\Z^{o}_2$ charge is confined and does not have any interplay with other symmetries.

The phase diagram of this dual model is shown in the second picture in Fig.~\ref{fig:SSBpert-2}. In the region $0\leq h<1$ and $0\leq p<\frac{1}{2}$, the mixed-state belongs to the even ANISPT phase because when $h=p=0$ the ground state of the Hamiltonian realizes the pure-state even NISPT \cite{Cao:2025qhg}.   The correlation function \eqref{eq:corr-2} is mapped to the string order parameter:
\begin{widetext}
\[\label{eq: even string}
\begin{split}
    &\text{Tr}\left[\rho \sigma^z_{4i-2}(1+\sigma^x_{4i-1})\sigma^y_{4i}\left(\prod^{4j-3}_{k=4i+1}\sigma^x_k\right)\sigma^y_{4j-2}(1+\sigma^x_{4j-1})\sigma^z_{4j}\right]\\
    =&\text{Tr}\left[\rho \sigma^z_{4i-2}(1+\sigma^x_{4i-1})\sigma^y_{4i}\left(\prod^{2j-2}_{k=2i+1}\sigma^x_{2k}\right)\sigma^y_{4j-2}(1+\sigma^x_{4j-1})\sigma^z_{4j}\right]\ne 0,
\end{split}
\]
\end{widetext}
where $\rho=\mathcal{N}[\rho'_0]$ and we use the fact $\sigma^x_{2i-1}\mathcal{N}[\rho'_0]=\mathcal{N}[\rho'_0]$.
This string operator is the disorder operator of even-site spin flip attached with boundary operators $\sigma^z_{4i-2}(1+\sigma^x_{4i-1})\sigma^y_{4i}$ and $\sigma^y_{4j-2}(1+\sigma^x_{4j-1})\sigma^z_{4j}$, which are both odd under $\mathsf{W}_1$ in \eqref{eq: W1 trans}. This string order characterizes the even ANISPT order and is consistent with the anticommutation relation between the actions of $U_e$ and $\mathsf{W}_1$ on the interface between pure-state even and trivial NISPTs in the reference \cite{Seifnashri:2024dsd}.

When $p=\frac{1}{2}$ and $0\leq h<1$, the mixed state belongs to $\Z^U_2$ SWSSB phase. In particular, the mixed state at $p=\frac{1}{2}$ and $h=0$ is obtained by applying KW duality on the mixed state \eqref{eq:SWSSB-1} and is given by
    \begin{equation}
    \rho=\frac{\mathbb{I}+\prod {\sigma}^x_i}{2}\prod_i \frac{\mathbb{I}+\sigma^x_{4i-3}}{2}\frac{\mathbb{I}+\sigma^z_{4i-4}\sigma^x_{4i-2}\sigma^z_{4i}}{2}\frac{\mathbb{I}+\sigma^x_{4i-1}}{2}.
\end{equation}
The first part of the above density matrix characterizes the $\Z^U_2$ SWSSB feature.

Moreover, when $p=1/2$, the mixed state after gauging can be characterized by the dual order parameter
\[
\text{Tr}(\rho \sigma^z_{4i}\sigma^z_{4j-1})
\]
and 
the dual Wightman disorder parameter:
\[
\text{Tr}(\sqrt{\rho} \mu_{i,j}\sqrt{\rho} \mu_{i,j}),
\]
where
\[
\mu_{i,j}=\sigma^z_{4i-2}(1+\sigma^x_{4i-1})\sigma^y_{4i}\left(\prod^{4j-2}_{k=2i+1}\sigma^x_{2k}\right)\sigma^y_{4j-2}(1+\sigma^x_{4j-1})\sigma^z_{4j}.\nonumber
\]
These two quantities follow from applying the KW duality on \eqref{eq: disorder-2} and \eqref{eq: weightman-2}.
When $0\leq h<1$, the dual order parameter exhibits exponential decay and the dual Wightman disorder parameter is long range in the $\Z^U_2$ SWSSB phase. When $h>1$, the dual order parameter is long range and the Wightman correlation function exhibits exponential decay in the $\Z^U_2$-SSB phase.  At mixed-state phase transition $h=1$, both two parameter have a power-law behavior $\frac{1}{|i-j|^{\frac{1}{4}}}$.

Finally, if we translate the model \eqref{eq: even HAL} and \eqref{eq: even channel} by one-site, the resulting model should have the same phase diagram as Fig.~\ref{fig:SSBpert-2} except replacing the even ANISPT by the odd ANISPT phases \cite{Seifnashri:2024dsd}. By one-site translation, the string order in \eqref{eq: even string}  becomes
\begin{widetext}
\[
\text{Tr}\left[\rho \sigma^z_{4i-1}(1+\sigma^x_{4i})\sigma^y_{4i+1}\left(\prod^{4j-2}_{k=2i+1}\sigma^x_{2k+1}\right)\sigma^y_{4j-1}(1+\sigma^x_{4j})\sigma^z_{4j+1}\right]\ne 0,
\]
\end{widetext}
when $0 \leq h<1$ and $0\leq p<1/2$.
Such string order is the disorder parameter of odd-site spin-flip attached with two local operators with odd $\mathsf{W}_1$ charge, which characterizes the odd ANISPT order.

\section{Lattice models of Rep$(D_{8})$ IANISPT phase}
In this section, we will construct the  fixed point wavefunction for the Rep($D_8$) IANISPT and then study the projective representation of Rep$(D_8)$ symmetry at the interface between this IANISPT and the trivial ASPT phase discussed in Sec. \ref{sec: trivial}.

The generalization to general Rep$(D_{4n})$ IANISPTs is straightforward and we leave it in App.~\ref{app:ianispt}.

\subsection{Fixed-point wavefunction}
We also begin with $D_8$ symmetric side, where the system is in the $\Z^U_2$ SSB phase with unbroken group $H=\Z^{\eta}_4$ and two ground states are both in $\Z^{\eta}_4$  IASPT. We group four qubits in one unit cell and consider the subspace with the following constraints:
\begin{equation}\label{eq: const-1}
\sigma^z_{4j-3}\sigma^z_{4j+1}= 1, \quad
\sigma^z_{4j-1}\sigma^z_{4j}=1,\quad \forall j\,.
\end{equation}
The first constraint is between adjacent unit cells and set $\sigma_{4j-3}$ in the $\Z^U_2$ SSB phase. The second constraint is within a single unit cell, which makes $\sigma_{4j-1}$ and $\sigma_{4j}$ in the ferromagnetic configuration. Therefore, the effective degrees of freedom in the subspace in each unit cell are two qubits. We define the following effective Pauli operators:
\begin{equation}\label{eq: eff Paul1}
\begin{split}
&\tau^x_{j}=\sigma^x_{4j-1}\sigma^x_{4j}, \tau^z_{j}=\sigma^z_{4j-1}=\sigma^z_{4j},\\
&\mu^z_{j-\frac{1}{2}}=\sigma^z_{4j-2},\mu^x_{j-\frac{1}{2}}=\sigma^x_{4j-2}.
\end{split}
\end{equation}
The unbroken symmetry becomes effectively:
\begin{equation}
\begin{split}
\eta\to \prod_{j=1}^L\exp\left(\frac{\pi i}{4}(1-\mu^z_{j-\frac{1}{2}})\right)\tau^z_j\,,
\end{split}
\end{equation}
which is the same as that of the conventional $\Z_4$ IASPT~\cite{ma2025topological}.

We further consider the $\Z^{\eta}_4$ symmetric subspace by imposing the constraint
\[\label{eq: const-2}
(\sigma^x_{4j-5}\sigma^x_{4j-4})\ \sigma^z_{4j-2}\sigma^z_{4j-3}\ (\sigma^x_{4j-1}\sigma^x_{4j})=1\,,
\]
which is $\Z^{\eta}_4$ invariant in the subspace satisfying~$\eqref{eq: const-1}$.
Written in the effective operators, this constraint becomes
\[\label{eq:IASPT constraint}
\tau^x_{j-1}\mu^z_{j-\frac{1}{2}}\tau^x_{j}=\sigma^z_{4j-3}=
\begin{cases}
    +1 ,\quad \forall j\\
    -1,\quad \forall j
\end{cases}\,,
\]
where the sign depends on the configuration of the SSB ground state.
Therefore, the fixed-point wavefunction of $\Z^U_2$ SSB stacked with a $\Z^{\eta}_4$ IASPT is given by
\begin{widetext}
\begin{equation}  
\begin{split}
\rho=&|\uparrow\cdots\uparrow\rangle\langle\uparrow\cdots\uparrow|_{4j-3}\otimes \sum_{\{x_{j}\}}\bigotimes_{j}\left(|\tau^x_{j}=x_{j}\rangle\langle \tau^x_{j}=x_{j}|\otimes
|\mu^z_{j-\frac{1}{2}}=x_{j-1}x_{j}\rangle\langle \mu^z_{j-\frac{1}{2}}=x_{j-1}x_{j}|\right)\\&+|\downarrow\cdots\downarrow\rangle\langle\downarrow\cdots\downarrow|_{4j-3}\otimes \sum_{\{x_{j}\}}\bigotimes_{j}\left(|\tau^x_{n}=x_{j}\rangle\langle \tau^x_{j}=x_{j}|\otimes
|\mu^z_{j-\frac{1}{2}}=-x_{j-1}x_{j}\rangle\langle \mu^z_{j-\frac{1}{2}}=-x_{j-1}x_{j}|\right)\,,
\end{split}
\end{equation}
\end{widetext}
where IASPT part is the maximally mixed state in the subspace \eqref{eq:IASPT constraint} and is attached to each ground state of the $\mathbb Z_2^{U}$-SSB phase on $\sigma_{4j-3}$.

Now let us gauging $\Z^U_2$ symmetry, i.e.~applying KW duality to constraints~\eqref{eq: const-1} and~\eqref{eq: const-2}, the corresponding subspace satisfying the dual constraints:
\begin{equation}\label{eq:noninv subspace}
\begin{split}
    & \sigma^x_{4j-1}=1, \quad \sigma^x_{4j-3}\sigma^x_{4j-2}\sigma^x_{4j}=1, \\
     &(\sigma^z_{4j-6}\sigma^z_{4j-4})\ \sigma^x_{4j-3}\ (\sigma^z_{4j-2}\sigma^z_{4j})=1.
\end{split}
\end{equation}
 Thus the Rep($D_8$)  IANISPT fixed-point wavefunction is the maximally mixed state in this subspace. 

To better describe the IANISPT mixed state, we define the effective operators  
\begin{align}
    \begin{split}\label{eq: eff Paul2}
   &\tau^x_{j}= \sigma^z_{4j-2}\sigma^z_{4j},\quad  \tau^z_{j}=\sigma^x_{4j},\\
   &\mu^z_{j+\frac{1}{2}}=\sigma^x_{4j+1},\quad  \mu^x_{j+\frac{1}{2}}=\sigma^z_{4j+1}\sigma^z_{4j+2},
    \end{split}
\end{align}
for the remaining degrees of freedom in each unit cell after imposing the first line of ~\eqref{eq:noninv subspace}. Note that the number of degrees of freedom per unit cell remains two qubits, as this is preserved under gauging.
These effective Pauli operators commute with the second constraint of~\eqref{eq:noninv subspace}.
With this notation, the last constraint in \eqref{eq:noninv subspace} is effectively  $\tau^x_{j-1}\mu^z_{j-\frac{1}{2}}\tau^x_{j}=1$. Hence the  Rep($D_8$)  IANISPT fixed-point state can written as
\begin{widetext}
\[\label{eq: IANISPT state}
\begin{split}
\rho_{\text{IANISPT}}=&\bigotimes_{j}\left(|\sigma^x_{4j-1}=1\rangle\langle\sigma^x_{4j-1}=1|\otimes |\sigma^x_{4j-3}\sigma^x_{4j-2}\sigma^x_{4j}=1\rangle\langle \sigma^x_{4j-3}\sigma^x_{4j-2}\sigma^x_{4j}=1|\right)\\&\otimes \sum_{\{x_{j}\}}\bigotimes_{j}\left(|\tau^x_{j}=x_{j}\rangle\langle \tau^x_{j}=x_{j}|\otimes
|\mu^z_{j-\frac{1}{2}}=x_{j-1}x_{j}\rangle\langle \mu^z_{j-\frac{1}{2}}=x_{j-1}x_{j}|\right)\,,
\end{split}
\]
\end{widetext}
realizing the maximally mixed state in the subspace \eqref{eq:noninv subspace}.

Moreover, such a mixed state in the subspace \eqref{eq:noninv subspace} can be identified by another set of constraints in the effective description:
\[\label{eq: mixed state const}
\begin{split}
\tau^x_{j-1}\mu^z_{j-\frac{1}{2}}\tau^x_{j}\rho&=\rho, 
\\
\tau^x_j \rho \tau^x_j&=\rho,  
\\
\mu^x_{j-\frac{1}{2}}\tau^z_j \mu^x_{j+\frac{1}{2}}\rho \mu^x_{j-\frac{1}{2}}\tau^z_j \mu^x_{j+\frac{1}{2}}&=\rho, \\
\mu^y_{j-\frac{1}{2}}\tau^z_j \mu^y_{j+\frac{1}{2}}\rho \mu^y_{j-\frac{1}{2}}\tau^z_j \mu^y_{j+\frac{1}{2}}&=\rho. 
\end{split}
\]
Here the first constraint ensures that $\rho$ is expanded in the subspace satisfying $\mu^z_{j-\frac{1}{2}}\tau^x_{j-1}\tau^x_{j}=1$ and the second constraint ensures that $\rho$ is the diagonal in $\tau^x$ basis. The last two constraints proliferate distinct $\tau^x$ configurations. Because the $\tau$ and $\mu$ part of~\eqref{eq: IANISPT state} realizes a $\Z^{\eta}_4$ IASPT in \cite{ma2025topological},  these above constraints indeed are the same as those of the $\Z_4$ IASPT mixed state in \cite{PRXQuantum.6.010347} if we do $\pi/2$ spin rotation on $y$ direction.

\subsection{The projective representation of symmetry on the interface}\label{sec: interface}
One signature of ASPT phases is the presence of edge modes, which comes from the projective representation of the symmetries at the interfaces between two distinct ASPTs.

To see the projective representation on the interfaces between this IANISPT and other ANISPTs, we first work out how the noninvertible symmetry $\mathsf{W}_{1}$ acts on the effective operators \eqref{eq: eff Paul2}.  Due to \eqref{eq: W1 trans}, we have the action of Rep$(D_8)$  on the effective operators as:
\begin{equation}
\begin{split}
 \mathsf{W}_1:   &~\tau^x_j \to -\tau^x_j, \tau^z_j \to \tau^z_j, \mu^z_{j+\frac{1}{2}}\to \mu^z_{j+\frac{1}{2}}, \\ &\mu^x_{j-\frac{1}{2}}\mu^x_{j+\frac{1}{2}}\to \mu^y_{j-\frac{1}{2}}\mu^y_{j+\frac{1}{2}},
 \end{split}
\end{equation}
and $ U_e$ and $U_o$ are both effectively $\prod_{j} \mu^z_{j+1/2}$.
Since $\mathsf{W}_{1}$ commutes with all $\tau^z_n$ and $\mu^z_n$, $W_1$ should be diagonal in $Z$-basis.  Moreover, due to the fusion rule:
\begin{equation}
    \mathsf{W}_{1}^2=(1+\prod_j \sigma^x_{2j})(1+\prod_j \sigma^x_{2j-1}),
\end{equation}
we have the following fusion rule in the subspace:
\begin{equation}
    \mathsf{W}_{1}^2=(1+\prod_j \sigma^x_{4j+1})(1+\prod_j \sigma^x_{4j+1})=2(1+\prod_j \mu^z_{j+\frac{1}{2}}),\nonumber
\end{equation}
where we use the first and second constraint in eq.~\eqref{eq:noninv subspace}  in the first equation.

Hence, the $\mathsf{W}_{1}$ must act effectively as:
\begin{equation}
    \mathsf{W}_{1}=(1+\prod_j \mu^z_{j+\frac{1}{2}})M,
\end{equation}
where $M$ is a diagonal matrix in $Z$-basis and it should be expanded by the operators $\tau^z$ and $\sigma^z$.
Moreover, as $\mathsf{W}_{1}$ anticommutes with $\tau^x_j$, the $\tau^z$ part of $M$ must be $\prod_j \tau^z_j$. Besides, since $ \mathsf{W}_{1}$ connects $\mu^x_{j-\frac{1}{2}}\mu^x_{j+\frac{1}{2}}$ and $ \mu^y_{j-\frac{1}{2}}\mu^y_{j+\frac{1}{2}} $, the $\mu$ part of $M$ should be $\prod_{j=1}^L\exp\left(\frac{\pi i}{4}(1-\mu^z_{j-\frac{1}{2}})\right)$. Thus in this subspace, we have
\begin{equation}\label{eq: eff sym}
    \mathsf{W}_{1}=(1+\prod_j \mu^z_{j+\frac{1}{2}})\prod_{j}\tau^z_{j}\prod_{j=1}^L\exp\left(\frac{\pi i}{4}(1-\mu^z_{j-\frac{1}{2}})\right).
\end{equation}


Now let us consider the following constraints in the effective
description:
\begin{equation}\label{eq: interface}
\begin{alignedat}{2}
\mu^z_{j-\frac{1}{2}}\tau^x_{j-1}\tau^x_{j}\,\rho &= \rho,
\quad & l+1\leq j\leq l',\\
\tau^x_j\, \rho \,\tau^x_j &= \rho,
\quad & l\leq j\leq l',\\
\mu^x_{j-\frac{1}{2}}\tau^z_j \mu^x_{j+\frac{1}{2}}\rho \mu^x_{j-\frac{1}{2}}\tau^z_j \mu^x_{j+\frac{1}{2}} &= \rho,
\quad & l\leq j\leq l'-1,\\
\mu^y_{j-\frac{1}{2}}\tau^z_j \mu^y_{j+\frac{1}{2}}\rho \mu^y_{j-\frac{1}{2}}\tau^z_j \mu^y_{j+\frac{1}{2}} &= \rho,
\quad & l\leq j\leq l'-1,\\
\mu^z_{j-\frac{1}{2}}=\tau^z_j &= 1,
\quad & j<l \ \text{or}\ j>l',
\end{alignedat}
\end{equation}
which can set the system realize the IANISPT phase in the region $l\leq j \leq l'$ and realize the trivial ANISPT phase $\rho=\otimes_{j}|\sigma^x_j=1\rangle\langle\sigma^x_j=1|$ in other regions.
Note that the first four constraints are the truncation of eq.~\eqref{eq: mixed state const}, which realizes the IANISPT mixed state in the region $[l,l']$. The last constraint gives the trivial ANISPT mixed state $\rho=\otimes_{j}|\sigma^x_j=1\rangle\langle\sigma^x_j=1|$ in the other regions. We denote the space of the corresponding mixed states as $\mathcal{C}$.

In the subspace represented by $\mu$ and $\tau$, the above interface is equivalent to the
interface between a $\Z_4$ IASPT and the trivial ASPT. This has been discussed in reference \cite{PRXQuantum.6.010347}, which only differs by a $\pi/2$ spin rotation on the $y$ direction. Here $\mathcal{C}$ is  4-dimensional which comes from edge modes at two interfaces. The effective Pauli operator of the edge modes are given by $\tilde\tau^x_{l'}=\tau^x_{l'},\tilde\tau^z_{l'}=\tau^z_{l'}\mu^x_{l'-\frac{1}{2}},\tilde\mu^x_{l-\frac{1}{2}}=\mu^x_{l-\frac{1}{2}},\tilde\mu^z_{l-\frac{1}{2}}=\tau^x_{l}\mu^z_{l-\frac{1}{2}}$. According to \cite{PRXQuantum.6.010347}, we obtain the symmetry transformation in the subspace $\mathcal{C}$ as
\begin{widetext}
\[
\mathsf{W}_{1}\rho_j \mathsf{W}_{1}=(1+\tau^x_{l}\mu^z_{l-\frac{1}{2}}\tau^x_{l'})\mu^x_{l-\frac{1}{2}}\mu^x_{l'-\frac{1}{2}}\tau^z_{l'}\rho_j (1+\tau^x_{l}\mu^z_{l-\frac{1}{2}}\tau^x_{l'})\mu^x_{l-\frac{1}{2}}\mu^x_{l'-\frac{1}{2}}\tau^z_{l'},\quad 
\prod_n \mu^z_{n-\frac{1}{2}}\rho_j=\mu^z_{l-\frac{1}{2}}\tau^x_l\tau^x_{l'}\rho_j. \quad \forall \rho_j\in \mathcal{C}.
\]
\end{widetext}
Thus the symmetry factorizes at the interface as
\begin{align}
\begin{split}
\mathsf{W}_{1}&=\mathscr{L}^1_{\mathsf{W}_{1}}\circ\mathscr{R}_{\mathsf{W}_{1}}^1+\mathscr{L}^2_{\mathsf{W}_{1}}\circ\mathscr{R}^2_{\mathsf{W}_{1}},\\
\prod_n\mu^z_n&=\mathscr{L}_{\mu}\circ\mathscr{R}_{\mu},
\end{split}
\end{align}
where
\begin{equation}
\begin{split}
&\mathscr{L}^1_{\mathsf{W}_{1}}=\tilde\mu^x_{l-\frac{1}{2}},\quad \mathscr{L}^2_{\mathsf{W}_{1}}=-\tilde\mu^y_{l-\frac{1}{2}},\quad
\mathscr{R}^1_{\mathsf{W}_{1}}=\tilde\tau^z_{l'},\\ &\mathscr{R}^2_{\mathsf{W}_{1}}=\tilde\tau^y_{l'},\quad \mathscr{L}_{\mu}=\tilde\mu^z_{l-\frac{1}{2}},\quad \mathscr{R}_{\mu}=\tilde\tau^x_{l'}.
\end{split}
\end{equation}
 Since $ U_e$ and $U_o$ both effectively correspond to $\prod_{j} \mu^z_{j+1/2}$, $\mathscr{L}_{\mu}/\mathscr{R}_{\mu}$ characterize the action of $U_e$ and $U_o$ on the interface modes. Moreover, due to $\mathscr{L}^{1/2}_{\mathsf{W}_{1}}/\mathscr{R}^{1/2}_{\mathsf{W}_{1}}$ anticommutes with  $\mathscr{L}_{\mu}/\mathscr{R}_{\mu}$,  the action of $\mathsf{W}_{1}$ anticommutes with the action of $U_e$ and $U_o$ on each interface. This projective representation is different from those on the interfaces between odd/even and trivial pure-state NISPTs in the reference \cite{Seifnashri:2024dsd}. This implies that the SPT order of this mixed-state phase is intrinsic average.


\section{Classification of $(1+1)d$ $\mathbb Z_N$ dipole ASPT}
In this section, we turn to another generalized symmetry -- dipole symmetry in $(1+1)d$. We first study how $\mathbb Z_N$ dipole symmetry is realized in the open quantum system. Then we use the gauging method to give a complete classification of dipole ASPT and IASPT.

\subsection{Dipole symmetry in open quantum systems}
As reviewed in Sec.~\ref{sec:dipolesym}, dipole symmetry can be generated by gauging $\Z^{X}_N$ subgroup of $\Z^X_{N}\times \Z^Z_N$ symmetry in the closed quantum systems. This gauging construction can be directly generalized to open quantum systems. Since we gauge the  $\Z^{X}_N$ subgroup, it must be part of the strong symmetry group. Moreover, we further require a subgroup of $\Z^{Z}_N$ to remain strong, while the quotient group to be weak. More precisely, for a given decomposition $N=pq$, we choose the subgroup $\Z^{Z^p}_q$ generated by $U_Z^p$ to be strong. Then after gauging $\Z^X_N$ symmetry, the image of the strong subgroup $\Z^{Z^p}_q$ is $\Z^{D^p}_q$ generated by $U_D^p$, which remains strong, and the corresponding quotient group is weak. Then a symmetric mixed state satisfies
\[\label{eq: symmetry conditions-2}
U_Q\rho=\rho,\quad U^p_D\rho=\rho, \quad
U_D\rho=\rho U_D. 
\]

\subsection{Classification of dipole ASPT and IASPT}
\label{sec: class of dASPT}
We turn to the classification of dipole ASPT and IASPT. As discussed in Sec. \ref{sec: gauging map}, gauging implements a one-to-one mapping between
mixed-state phases. Hence the above problem is equivalent to the classification of $\Z^X_N$ SSB phase in $\Z^{X}_N\times \Z^Z_N\times \Z^{\text{Tr}}$ symmetric systems before gauging. Here $\Z^{\text{Tr}}$ is the translation symmetry. There are $N$ distinct choices of unbroken group $H_{m}=\Z^{ZX^m}_N$ which is generated by $U_Z U^m_X$ with $m=0,1,\cdots , N-1$. For each $H_m$, the strong subgroup is $A_m=H_m\cap K=\mathbb{Z}_{q}^{Z^p X^{pm}}$, which is generated by $U^p_Z U^{pm}_X$, and the corresponding quotient $\Z_p$ group is denoted by $B_m=H_m/K_m$. Since all the unbroken group $H_m$ commutes with translation, these SSB phases are all invariant under translation.

For the dipole ASPT phases, they correspond to the $\Z^X_N$ SSB phases stacking with the $H_m$-ASPT phases. Since $H^2(H_m, U(1))=\Z_1$, there is only a trivial $H_m$-ASPT for each choice of unbroken group. Therefore, there are $N$ distinct $\Z^X_N$ SSB phases in this case, giving rise to $N$
distinct dipole ASPT phases respectively. This is consistent with the pure-state classification of dipole SPTs~\cite{Lam:2023xng}.

For the dipole IASPT phases, they arise from stacking 
$\Z^X_N$ SSB phases with a $H_m$-IASPT phase. For a given $H_m$, the IASPT phases can be classified by domain wall decoration~\cite{10.21468/SciPostPhys.17.1.013,PhysRevB.107.125158,ma2025topological} as follows:
\begin{enumerate}
    \item In the first step, let us denote the background fields of $A_m$ and $B_m$ as $a$ and $b$ respectively, both of which are 1-cochains. The corresponding $\Z_N$ background gauge field is $pa+b$. By requiring the $\Z_N$ background gauge field to be flat, we find
\begin{eqnarray}
\delta( pa + b)=0 \mod N,
\end{eqnarray}
which implies 
\begin{eqnarray}\label{eq: bundle}
\delta a=  -\frac{1}{p}\delta b \mod q, ~~~~~ \delta b=0\mod p.
\end{eqnarray}
Here $\delta$ is the differential operator on the cochain \cite{Wang:2021nrp}
\item In the second step, we decorate the weak symmetry domain wall with a strong charge. That is stacking a strong symmetry Wilson line to the worldline of a weak symmetry domain wall on a closed loop $[b]$ , which gives rise to a topological term \cite{PhysRevLett.114.031601}:
\begin{eqnarray}\label{eq:DDWano}
\exp\left(\frac{2\pi i k}{[p,q]} \int_{[b]} a\right) = \exp\left(\frac{2\pi i k}{[p,q]}  \int_{M_2} a\cup b\right).
\end{eqnarray}
Here the $[p,q]$ is the greatest common divisor of $p$ and $q$, and the level $k=0,1. \cdots, [p,q]-1$.
However, due to the nontrivial bundle constraint \eqref{eq: bundle}, the domain wall decoration is not gauge invariant, and equivalently it induces a nontrivial dependence on the extension to the 3d bulk $M_3$,
\begin{eqnarray}\label{eq: DDW}
\exp\left(\frac{2\pi i k}{[p,q]}  \int_{M_2} a\cup b\right) = \exp\left(-\frac{2\pi i k}{[p,q]p}\int_{M_3} \delta b\cup b\right).
\end{eqnarray}
In the second equality, we applied total derivative to promote the 2d integral to the 3d integral and used \eqref{eq: bundle}.
\item However, the IASPT should be independent of the extension to $M_3$. This demands that the system before domain wall decoration should already exhibit an opposite topological term of $B_M$, whose action is given by \cite{PhysRevLett.114.031601}
\begin{eqnarray}\label{eq: lowenergyanom}
\exp\left(\frac{2\pi i k'}{p^2}\int_{M_3} b\cup \delta b\right),
\end{eqnarray}
where the level $k'=0,1,\cdots p-1$.
After domain wall decoration, the  term \eqref{eq: lowenergyanom} should cancel against the induced term \eqref{eq: DDW} from the domain wall decoration, which implies $k'=kp/[p,q] $. As a result there will be $([p,q]-1)$ distinct IASPTs for a given $H_m$, where we exclude the trivial phase with $k=0$. These IASPTs have different representations of weak symmetries and decorated domain wall data.
\end{enumerate}

Since the classification of IASPT is the same for different $H_m$ branches, there are $N([p,q]-1)$ distinct $\Z^X_N$ SSB phase stacking with IASPT of the unbroken group. Therefore we obtain that there are in total $N([p,q]-1)$ distinct dipole IASPT phases. 

Here we remark that the above classification based on the background gauge field also works for the intrinsic gapless SPT (igSPT), which is shown in \cite{10.21468/SciPostPhys.17.1.013,PhysRevB.107.125158}. However, there is a difference between igSPT and IASPT. The 3d bulk term \eqref{eq: DDW}, \eqref{eq: lowenergyanom} is anomaly inflow action in such closed quantum systems. In particular, \eqref{eq: lowenergyanom} corresponds to  anomalous low-energy symmetries for igSPTs, which guarantee the long-range entanglement properties. On the other hand, weak symmetry with topological term \eqref{eq: lowenergyanom} is nonanomalous in the mixed-state setting and thus IASPT is still short-range entangled.

\section{Lattice models of dipole IASPT phases}

In this section, we turn to the lattice construction of dipole IASPT phases. We first give a general proof that the dipole IASPT in different branches is distinguished by stacking distinct dipole ASPT phases, which implements a unitary transformation realized as the decorated domain wall transformation. Then, we explicitly construct the fixed-point wavefunction for dipole IASPT with $N=4$ and $p=2$, together with the analysis of the projective representation at the interfaces.

\subsection{Unitary transformation between different $H_m$ branches}
For the $\Z^{X}_N$ SSB phases in the $\Z^{X}_N\times \Z^{Z}_N\times \Z^{\text{Tr}}$ symmetric system, there is a unitary operator $U_{H}$ which connects different unbroken group $U_{H}: H_{m} \to H_{m-1}$ and commutes with the $\Z^X_N$ and translation symmetry. Therefore, all the $\Z^{X}_N$ SSB phases can be constructed from those in the branch with unbroken group $H_0=\Z^Z_{N}$.  

For even $N$, this  unitary operator is given by
\[
U_{H}=\sum_{\{x_j\}}\omega^{\sum_{j}\frac{x^2_j}{2}}|\{x_j\}\rangle\langle\{x_j\}|,
\]
where $|x_j\rangle$ is an eigenstate of $X_j$ with eigenvalue $\omega^{x_j}$ and $x_j=0,\cdots, N-1$. This unitary transformation satisfies
\[\label{eq: unitary rela}
U_{H}U_Z U^{\dagger}_{H}=  U_ZU^{-1}_X, ~ U_{H}U_X=U_XU_{H},~ U_{H}T=TU_H,
\]
where $T$ is the translation operator.
Note that the phase $\omega^{\frac{x^2_j}{2}}$ is well defined for $\Z_N$ valued $x_j$ for even $N$ since $(x_j+N)^2=x^2_j \text{ (mod 2$N$)}$.

For odd $N$, this  unitary operator is given by
\[
U_{H}=\sum_{\{x_j\}}\omega^{\sum_{j}\frac{(1-N)x^2_j}{2}}|\{x_j\}\rangle\langle\{x_j\}|,
\]
which satisfies the same relation \eqref{eq: unitary rela}. In this case, the phase $\omega^{\sum_{j}\frac{(1-N)x^2_j}{2}}$ is well defined for $\Z_N$ valued $x_j$ since $(1-N)(x_j+N)^2=(1-N)x^2_j \text{ (mod 2$N$)}$.

Then after gauging the $\mathbb Z^X_N$ symmetry, equivalent to the generalized KW transformation~\eqref{eq:ZNKW},
the dual unitary transformation is given by
\[
\sum_{\{z_j\}}\omega^{\sum_{j}\frac{(z_{j-1}-z_{j})^2}{2}}|\{z_j\}\rangle\langle\{z_j\}|
\] for even $N$ and 
\[
\sum_{\{z_j\}}\omega^{\sum_{j}\frac{(1-N)(z_{j-1}-z_{j})^2}{2}}|\{z_j\}\rangle\langle\{z_j\}|
\]
for odd $N$. Here $|z_j\rangle$ is an eigenstate of $Z_j$ with eigenvalue $\omega^{z_j}$ and $z_j=0,\cdots, N-1$.

Under PBC boundary condition $z_{j+L}=z_j ~\text{mod}~ N$, this dual unitary transformation can be written in a unified presentation:
\[
U_{DDDW}=\sum_{\{z_j\}}\omega^{\sum_{j}(-z_{j-1}z_{j}+z^2_j)}|\{z_j\}\rangle\langle\{z_j\}|,
\]
which is the decorated domain wall transformation connecting distinct pure-state dipole SPT phases~\cite{Han:2023fas}. Therefore, the dipole IASPT phases constructed from different unbroken group branches are distinguished by stacking distinct dipole SPT phases.
\subsection{Lattice model for dipole IASPT with $N=4$ and $p=2$}
In this subsection, we construct a fixed-point wavefunction for the dipole IASPT phases with $N=4$ and $p=2$. There are $4$ distinct dipole IASPT phases. According to the previous section, we will focus on the IASPT phase corresponding to the unbroken group $H_0=\Z^Z_4$ before gauging and other IASPT phases can be obtained from this phase by $U_{DDDW}$ transformation. Since $p=2$, the strong symmetry is $A_0=\Z^{Z^2}_2$ and the weak $\Z_2$ symmetry is generated by $U_Z$. 

Let us start with $\Z^X_4$ SSB side. We consider a $\Z_4$ qudit ladder and the symmetry is generated by
\[
U_X=\prod^L_{j=1}X_{j,1},\quad U_Z=\prod^L_{j=1}Z_{j,1}.
\]
where the $X_{j,1}$/ $X_{j,2}$ and $Z_{j,1}$/$Z_{j,2}$ is the operator on the $j$-th site on the first/second chain.

Then we consider the subspace with the following constraints:
\begin{equation}\label{eq:Z4X SSB}
Z^{\dagger}_{j-1,1}Z_{j-1,2}Z_{j,1}Z^{\dagger}_{j,2}=1,
\end{equation}
where we group the same site in the first and second chains in one unit cell. This constraint gives the $\Z^X_4$ SSB phase with order parameter  $Z_{j,1}Z^{\dagger}_{j,2}$. We further note the effective degree of freedom (DOF) in this subspace \eqref{eq:Z4X SSB} is a single $\Z_4$ qudit in each unit cell in each SSB sector. Thus, we can decompose this effective DOF into two spin-1/2s with effective operators:
\[
\begin{split}
& \mu^z_{j-\frac{1}{2}}=Z^2_{j,1},\quad \mu^x_{j-\frac{1}{2}}=X_{j,1}X_{j,2} C_{j,1}(Z_{j,1})^{(1-X^2_{j,2})},\\ &  \exp\left(\frac{\pi i}{4}(1-\mu^z_{j-\frac{1}{2}})\right)\tau^z_j=Z_{j,1},\quad \tau^x_{j}=X^2_{j,1}X^2_{j,2}.
\end{split}
\]
Here the effective operators commute with \eqref{eq:Z4X SSB} and $C_{j,1}$ is the charge conjugation operator on $j$-th site on the first chain. Then the $\Z^Z_4$ symmetry can be written as
\[
U_Z\to \prod_{j=1}^L\exp\left(\frac{\pi i}{4}(1-\mu^z_{j-\frac{1}{2}})\right)\tau^z_j.
\]

Next we further consider the constraint:
\[\label{eq:Z4ZIASPT}
Z^2_{j,2}X^2_{j-1,1}X^2_{j-1,2}X^2_{j,1}X^2_{j,2}=1,
\]
which commutes with the constraint in \eqref{eq:Z4X SSB} and the $\Z^X_4\times \Z^Z_4\times \Z^{\text{Tr}}$ symmetry. In each SSB phase sector, i.e. $Z_{j,1}Z^{\dagger}_{j,2}=i^{k}$ where $k=0,1,2,3$, the above constraint is equivalent to  
\[
(-1)^k Z^2_{j,1}X^2_{j-1,1}X^2_{j-1,2}X^2_{j,1}X^2_{j,2}=1.
\]
Written in the effective operators, this above constraint further becomes $(-1)^k\mu^z_{j-\frac{1}{2}}\tau^x_{j-1}\tau^x_j=1$
which is the same as that of the conventional $\Z_4$ IASPT.

Therefore, the fixed-point wavefunction of $\Z^{Z}_4$ IASPT with $\Z^X_4$ SSB is given by
\begin{widetext}
\begin{equation}  
\begin{split}
\rho_{\text{IASPT}}=&\sum^3_{k=0}\bigg(\bigotimes_j| Z_{j,2}=i^k Z_{j,1}\rangle\langle Z_{1,j}=i^k Z_{2,j}|_j\\
&\otimes \sum_{\{x_{j}=\pm 1\}}\bigotimes_{j}\left(|\tau^x_{j}=x_{j}\rangle\langle \tau^x_{j}=x_{j}|\otimes
|\mu^z_{j-\frac{1}{2}}=(-1)^{k}x_{j-1}x_{j}\rangle\langle \mu^z_{j-\frac{1}{2}}=(-1)^k x_{j-1}x_{j}|\right)\bigg)\,,
\end{split}
\end{equation}
\end{widetext}
where IASPT part is the maximally mixed state in the subspace \eqref{eq:Z4ZIASPT}.

Now let us gauge $\Z^X_4$ symmetry, the dual system
has $\mathbb Z_N^Q\times \mathbb Z_N^D$ charge and dipole symmetry:
\begin{equation}\label{eq:dipolesymapp}
    U_Q=\prod_{j=1}^LX_{j,1},\quad U_D=\prod_{j=1}^L(X_{j,1})^j,
\end{equation}
which only acts on the first chain, and the translation symmetry.
The corresponding subspace satisfies the constraint :
\begin{equation}\label{eq: dipole const}
Z^{\dagger}_{j,2}Z_{j+1,2}X^{\dagger}_{j,1}=1, \quad
Z^2_{j,2}X^2_{j-1,2}X^2_{j,2}Z^2_{j-2,1}Z^2_{j,1}=1.
\end{equation}
which is obtained by applying KW duality to \eqref{eq:Z4X SSB} and \eqref{eq:Z4ZIASPT}.
To further describe the fixed-point wavefunction of IASPT, we define the effective operators, which commute with the first constraint in \eqref{eq: dipole const}: 
\begin{equation}
\begin{split}
     \exp\left(\frac{\pi i}{4}(1-\mu^z_{j-\frac{1}{2}})\right)\tau^z_j=Z_{j,2},\quad \tau^x_{j}=Z^2_{j-1,1}Z^2_{j,1}X^2_{j,2}.\nonumber
    \end{split}
\end{equation}
Here the number of degrees of freedom per unit cell remains two qubits, as this is preserved under gauging.

By this notation, the second constraint in \eqref{eq: dipole const} becomes effectively  $\mu^z_{j-\frac{1}{2}}\tau^x_{j-1}\tau^x_{j}=1$. Hence the  $\Z_4$ dipole symmetry acts effectively
\[
U_D \to \prod^L_{j=1}Z^{\dagger}_{j,2}\to \prod_{j=1}^L\exp\left(-\frac{\pi i}{4}(1-\mu^z_{j-\frac{1}{2}})\right)\tau^z_j
\]
and the IASPT fixed-point wavefunction can be written as
\begin{widetext}
\begin{equation}  
\begin{split}
\rho=\bigotimes_j| X_{1,j}= Z^{\dagger}_{j,2}Z_{j+1,2}\rangle\langle X_{1,j}= Z^{\dagger}_{j,2}Z_{j+1,2}|_j\otimes \sum_{\{x_{j}\}}\bigotimes_{j}|\tau^x_{j}=x_{j}\rangle\langle \tau^x_{j}=x_{j}|\otimes
|\mu^z_{j-\frac{1}{2}}=x_{j-1}x_{j}\rangle\langle \mu^z_{j-\frac{1}{2}}= x_{j-1}x_{j}|.
\end{split}
\end{equation}
\end{widetext}
This wavefunction is the maximally mixed state in the subspace \eqref{eq: dipole const}.

Now let us consider the interface between this dipole IASPT and the trivial ASPT phase. More precisely, system realizes the IASPT phase in the region $[l,l']$ and realizes the trivial ASPT phase $\rho=\otimes_{j}|X_{j,1}=Z_{j,2}=1\rangle\langle X_{j,1}=Z_{j,2}=1|$, which corresponds to $\otimes_{j}|\tau^z_{j}=\mu^z_{j}=1\rangle\langle \tau^z_{j}=\mu^z_{j}=1|$ by the effective operators, in other regions. That is we consider the same constraint for the effective operators $\mu$ and $\tau$ as \eqref{eq: interface} in the subspace satisfying the first constraint of \eqref{eq: dipole const}. Therefore there are also 4-fold mixed states which correspond to edge modes at the interface and the effective Pauli operator of the edge modes are given by $\tilde\tau^x_{l'}=\tau^x_{l'},\tilde\tau^z_{l'}=\tau^z_{l'}\mu^x_{l'-\frac{1}{2}},\tilde\mu^x_{l-\frac{1}{2}}=\mu^x_{l-\frac{1}{2}},\tilde\mu^z_{l-\frac{1}{2}}=\tau^x_{l}\mu^z_{l-\frac{1}{2}}$.
Following the same calculation in \cite{PRXQuantum.6.010347}, the symmetry factorizes at the interface as
\[
\begin{split}
U_D&=\mathscr{L}_{D}\circ\mathscr{R}_{D},\quad
U_{D^2}=\mathscr{L}_{D^2}\circ\mathscr{R}_{D^2}
\\
\mathscr{L}_{D}&=\tilde\mu^x_{l-\frac{1}{2}},\quad
\mathscr{R}_{D}=\tilde\tau^z_{l'},\\ \mathscr{L}_{D^2}&=\tilde\mu^z_{l-\frac{1}{2}}, \quad
\mathscr{R}_{D^2}=\tilde\tau^x_{l}.
\end{split}
\]
This anticommuting relation between $U_D$ and $U^2_D$ on each interface is different from those on the interface between pure-state dipole SPTs discussed in~\cite{Han:2023fas}. Thus this implies the SPT order of this mixed state is the intrinsic average.

\section{Conclusion}
We introduced the gauging method to study ASPT phases with generalized symmetries, focusing on noninvertible and dipole symmetry in $(1+1)d$. We prove that gauging defines a one-to-one correspondence between mixed-state phases on the two sides under the two-way connectivity definition. Using this framework, we systematically classify both nonintrinsic and intrinsic ASPTs with mixed strong–weak noninvertible Rep$(D_{4n})$ symmetry and $\mathbb Z_N$ dipole symmetry. Moreover, our approach enables direct lattice constructions of these ASPT phases. Since gauging can be implemented as quantum operations \cite{PhysRevX.14.021040}, our results also provide a concrete recipe for the operational preparation of these ASPT phases from mixed-state phases with conventional group symmetries, which we leave for future study.

Furthermore, the gauging method has proven powerful in the classification and construction of noninvertible-symmetry-protected topological (NISPT) phases in closed systems beyond $(1+1)d$. It is therefore natural to extend our analysis to ANISPTs in higher dimensions. Here we illustrate the method in the simplest setting of gauging a $\mathbb{Z}_N$ symmetry. It would be interesting to extend the construction to more general gauging procedures and thereby access additional topological phases, such as ASPTs with $G \times \mathrm{Rep}(G)$ symmetries. Finally, this gauging approach can apply to more general mixed-state phases beyond ASPTs, as we demonstrate via a one-to-one correspondence between mixed-state phases on the two sides. This motivates further applications of this method to spontaneous symmetry breaking phases and symmetry-enriched topological phases with generalized symmetries in the open systems.

\begin{acknowledgements}
{\it Acknowledgments}.---W.C.~acknowledges the support from Villum Fonden Grant no.~VIL60714. L.H.L. and Z.B. acknowledge a startup fund and a Quantum SuperSEED fund from the Pennsylvania State University. ZB also acknowledges support from NSF under award number DMR-2339319.
\end{acknowledgements}

\bibliography{bib}

\clearpage
\appendix
\section{Classification of $\mathbb Z_{n}^{\eta^2}\rtimes\mathbb Z_2^{\eta U}$-ASPT}\label{app1}
In this appendix, we will discuss the  classification of $\mathbb Z_{n}^{\eta^2}\rtimes\mathbb Z_2^{\eta U}$ ASPT phases with strong symmetry $K=\Z_{n}^{\eta^2}$ and the weak symmetry $G=\Z_2^{\eta U}$. 

According to \cite{ma2025topological}, all ASPT phases protected can be constructed using the decorated domain-wall construction. The idea is as follows:
\begin{enumerate}
    \item One decorates the domain walls of the weak symmetry with charges of the strong symmetry.
    \item If the decoration is consistent, i.e. the corresponding cochain satisfies the cocycle condition by itself, then the resulting phase is a non-intrinsic ASPT.
    \item If the decoration is inconsistent, but the obstruction can be canceled by a one higher dimensional decoration, then the resulting phase is an IASPT.
\end{enumerate}

In our case, the possible decorations are classified by $H^1[\Z_2, H^1[\Z_n,U(1)]]=\Z_{[n,2]}$. But it is known that $H^2[D_{2n},U(1)]=\Z_{[n,2]}$ and $H^2[\Z_{n},U(1)]=\Z_{1}$, thus  the classification of non-intrinsic ASPT phases, which is classified by  $H^2[D_{2n},U(1)]/H^2[\Z_{n},U(1)]$,  is the same as that of possible decorations. This implies that all allowed decorations lift to non-intrinsic ASPTs and no decoration produces an obstruction. In other words, all decorations are cohomologically consistent and therefore realize non-intrinsic ASPT phases only.  When $n$ is odd there is one ASPT phase, while when $n$ is even  there are two ASPT phases.

\section{General Rep($D_{4n})$ ANISPT Lattice models}\label{app:anispt}
In this appendix, we use the gauging-based framework to construct non-intrinsic Rep($D_{4n})$ ANISPT phases. We will start from $D_{4n}$ symmetry and gauge the non-normal $\mathbb Z_2^{U}$ to get Rep($D_{4n})$ symmetry. We work on spin chain with $L$ sites with $L=0~\text{mod}~2n$ and group $2n$ sites in one unit cell.

To realize the trivial Rep($D_{4n})$ ANISPT phase, we start from the following $D_{4n}$ symmetric Hamiltonian
\begin{equation}
    H=-\sum_{j=1}^L\sigma_j^{z}\sigma_{j+1}^{z}-\frac{h}{2n}\sum_{i=1}^{L/(2n)}\sum_{l=1}^{2n}\eta^{l}\left(\prod_{i_0=0}^{2n-1}\sigma_{2n(i-1)+i_0}^{x}\right)\eta^{-l},
\end{equation}
with the following quantum channels acting on the ground state
\begin{align}\label{eq:quantumchannelapp}
    \begin{split}
      &\mathcal{N}[\rho]=\otimes_{i=1}^{L/(2n)}\mathcal{N}_i[\rho],\\
      &\mathcal{N}_i[\rho]=(1-p)\rho+\frac{p}{2}\sum_{k=1}^{2}e^{i\frac{\pi}{2n}O_i^{(k)}}\rho e^{i\frac{\pi}{2n}O_i^{(k)}},  
    \end{split}
\end{align}
where 
\begin{align}
    \begin{split}
       &O_i^{(1)}=\sum_{l=0}^{n-1}\eta^{2l}\left(\prod_{i_0=1}^{2n}\sigma_{2n(i-1)+i_0}^{x}\right)\eta^{-2l},\\
        &O_i^{(2)}=\eta O_i^{(2)} \eta^{-1}. 
    \end{split}
\end{align}
It is easy to check that this channel preserves $\eta$ weakly, but $\eta^2$ and $U$ strongly. Under the same logic as in the main text, the channel $\mathcal{N}[\rho_0]$ can be expanded in the subspace contained by $\sigma_{2n(i-1)+2j-1}^z\sigma_{2n(i-1)+2j}^z=1, \forall j=1,...,n$ in the $i$th unit cell. Therefore, the spins in a single unit cell are in the same direction and  we can go to the effective description within the $i$th unit cell
\begin{equation}
    |\tilde{\uparrow}\rangle_i=\otimes_{j=1}^{2n}\ket{\uparrow}_{2n(i-1)+j}, \quad  |\tilde{\downarrow}\rangle_i=\otimes_{j=1}^{2n}\ket{\downarrow}_{2n(i-1)+j}.\nonumber
\end{equation}
In this effective low energy subspace, we get the same effective Ising type Hamiltonian
\begin{equation}
    \tilde{H}=-\sum_{i=1}^{L/(2n)}\tilde{\sigma}_{i}^z\tilde{\sigma}_{i+1}^z-\tilde{\sigma}_{i}^x,
\end{equation}
together with the effective quantum channel
\begin{equation}
    \tilde{\mathcal{N}}_i[ \tilde{\rho}]=(1-p) \tilde{\rho}+p\tilde{\sigma}_i^{x} \tilde{\rho} \tilde{\sigma}_i^{x},
\end{equation}
as shown in the main text. We have the strong $\mathbb Z_2^U$ SSB phase when $0<h<1, 0<p<1/2$ and after KW transformation we get trivial Rep($D_{4n}$) ANISPT in the same region.

To realize the even Rep($D_{4n}$) ASPT, we consider the following $D_{4n}$ symmetric Hamiltonian
\begin{equation}
    H=H_0-\frac{h}{2n}\sum_{i=1}^{L/(2n)}\sum_{l=1}^{2n}\eta^{l}\left(\prod_{i_0=0}^{2n-1}\sigma_{2n(i-1)+i_0}^{x}\right)\eta^{-l},
\end{equation}
where $H_0$ is the Hamiltonian of even Rep($D_{4n}$) SPT~\cite{Cao:2025qhg}, with the same quantum channel~\eqref{eq:quantumchannelapp}. Same as in Sec.~\ref{sec:evenodd}, this channel can be expanded in the subspace constrained by the ground state condition of each cluster in $H_0$. Therefore, we can go to the effective description within one unit cell
\begin{align}
    \begin{split}
      |\tilde{\uparrow}\rangle_i&=\frac{1}{\sqrt{2}}(|\underbrace{\uparrow \, \cdots \, \uparrow}_{2n}\rangle+|\underbrace{\downarrow \, \cdots \, \downarrow}_{2n}\rangle), \\  
     |\tilde{\downarrow}\rangle_i&=\frac{1}{\sqrt{2}}(
     |\underbrace{\uparrow \, \cdots \, \uparrow}_{n}
\underbrace{\downarrow \, \cdots \, \downarrow}_{n}\rangle
-|\underbrace{\downarrow \, \cdots \, \downarrow}_{n}
\underbrace{\uparrow \, \cdots \, \uparrow}_{n}\rangle).  
    \end{split}
\end{align}
In this effective low energy subspace, we get the same effective Ising type Hamiltonian
\begin{equation}
    \tilde{H}=-\sum_{i=1}^{L/(2n)}\tilde{\sigma}_{i}^y\tilde{\sigma}_{i+1}^y-\tilde{\sigma}_{i}^z,
\end{equation}
together with the effective quantum channel
\begin{equation}
    \tilde{\mathcal{N}}_i[ \tilde{\rho}]=(1-p) \tilde{\rho}+p\tilde{\sigma}_i^{z} \tilde{\rho} \tilde{\sigma}_i^{z},
\end{equation}
as shown in the main text. We have the strong $\mathbb Z_2^U$ SSB phase when $0<h<1, 0<p<1/2$, and after KW transformation we get the even Rep($D_{4n}$) ASPT in the same region.

Furthermore, if $n$ is even, we can do one-site lattice translation on the above even ANISPT model, the resulting model should has the same phase diagram except replacing the even ANISPT by the odd ANISPT phases.

\section{General Rep($D_{4n})$ IANISPT mixed state}\label{app:ianispt}
Let us construct the fixed-point mixed-state realizing Rep$(D_{4n})$ IANISPT for even $n$. We again begin with $D_{4n}$ symmetric system whose unbroken symmetry is $\Z^{\eta}_{4n}$, but now we group $n'=3n/2+1$ spins in one unit cell.

We first restrict to the subspace defined by
\begin{equation}\label{eq: const-1 app}
\begin{split}
&\sigma^z_{n'(j-1)+1}\sigma^z_{n'j+1}= 1, \\
&\sigma^z_{n'(j-1)+k}\sigma^z_{n'(j-1)+k+1}=1,~ \text{when }1< k< n', k\ne n/2+1.
\end{split}
\end{equation}
The first constraint fixes the $\sigma^z_{n'(j-1)+1}=\pm 1$ corresponding to the the $\Z^U_2$ strong SSB phase. The second equation enforces that the spins $\sigma^z_{n'(j-1)+k}$ with $2\leq k\leq n/2+1$ form a ferromagnetic block and the spins $\sigma^z_{n'(j-1)+k}$ with $n/2+2\leq k\leq n'$ form another ferromagnetic block. Thus each unit cell contributes two effective qubits. We define the corresponding effective Pauli operators:
\begin{equation}\label{eq: eff Paul1}
\begin{split}
&\tau^x_{j}=\prod^{n'}_{k=n/2+2}\sigma^x_{n'(j-1)+k}, ~\tau^z_{j}=\sigma^z_{n'j},\\
&\mu^x_{j-\frac{1}{2}}=\prod^{n/2+1}_{k=2}\sigma^x_{n'(j-1)+k},\mu^z_{j-\frac{1}{2}}=\sigma^z_{n'(j-1)+2}.
\end{split}
\end{equation}
In this subspace, the unbroken symmetry acts effectively as
\begin{equation}
\begin{split}
\eta\to \prod_{j=1}^L\exp\left(\frac{\pi i}{4}(1-\mu^z_{j-\frac{1}{2}})\right)\tau^z_j,
\end{split}
\end{equation}
which matches the conventional $\Z_4$ IASPT action.

Next, we impose the additional constraint
\[\label{eq: const-2 app}
\sigma^z_{n'(j-1)+1}\sigma^z_{n'(j-1)+2}\prod^{n/2+1}_{k=2}\sigma^x_{n'(j-1)+k}\sigma^x_{n'j+k}=1,
\]
which is equivalent to 
\[\label{eq:IASPT constraint app}
\mu^z_{j-\frac{1}{2}}\tau^x_{j-1}\tau^x_{j}=\sigma^z_{n'(j-1)+1}=\pm 1.
\]
The sign corresponds to the $\Z^U_2$ SSB ground state. Note that the constraint \eqref{eq: const-2 app} is $\Z^{\eta}_{4n}$ invariant in the subspace $\eqref{eq: const-1 app}$.
Therefore, the fixed-point state of $\Z^{\eta}_{4n}$ IASPT with broken $\Z^U_2$ symmetry is therefore
\begin{widetext}
\begin{equation}  
\begin{split}
\rho&=|\uparrow\cdots\uparrow\rangle\langle\uparrow\cdots\uparrow|_{n'(j-1)+1}\otimes \sum_{\{x_{j}\}}\bigotimes_{j}|\tau^x_{j}=x_{j}\rangle\langle \tau^x_{j}=x_{j}|\otimes
|\mu^z_{j-\frac{1}{2}}=x_{j-1}x_{j}\rangle\langle \mu^z_{j-\frac{1}{2}}=x_{j-1}x_{j}|\\&+|\downarrow\cdots\downarrow\rangle\langle\downarrow\uparrow\cdots\downarrow|_{n'(j-1)+1}\otimes \sum_{\{x_{j}\}}\bigotimes_{j}|\tau^x_{n}=x_{j}\rangle\langle \tau^x_{j}=x_{j}|\otimes
|\mu^z_{j-\frac{1}{2}}=-x_{j-1}x_{j}\rangle\langle \mu^z_{j-\frac{1}{2}}=-x_{j-1}x_{j}|,
\end{split}
\end{equation}
\end{widetext}
where IASPT part is the maximally mixed state in the subspace \eqref{eq:IASPT constraint app}.

After gauging $\Z^U_2$ symmetry, the corresponding subspace satisfying the constraint:
\begin{equation}\label{eq:noninv subspace app}
\begin{split}
    & \sigma^x_{n'(j-1)+k}=1, ~ \text{when }1< k< n', k\ne n/2+1,
    \\
    &\sigma^x_{n'(j-1)+1}\sigma^x_{n'(j-1)+n/2+1}\sigma^x_{n'j}=1,\\
    &\sigma^x_{n'(j-1)+1}\sigma^z_{n'(j-1)+n/2+1}\sigma^z_{n'j}\sigma^z_{n'j+n/2+1}\sigma^z_{n'(j+1)}.
\end{split}
\end{equation} 

Due to \eqref{eq:noninv subspace app}, this subspace again contains two effective per unit cell. Defining effective operators that commute with the second constraint,
\begin{equation}\label{eq: eff Paul2 app}
\begin{split}
    &\tau^x_{j}=\sigma^z_{n'(j-1)+n/2+1}\sigma^z_{n'j}, \tau^z_{j}=\sigma^x_{n'j}, \\ &\mu^z_{j-\frac{1}{2}}=\sigma^x_{n'(j-1)+1}, \mu^x_{j-\frac{1}{2}}=\sigma^z_{n'(j-1)+1}\sigma^z_{n'(j-1)+n/2+1}.
    \end{split}
\end{equation}
Then the last constraint in \eqref{eq:noninv subspace app} becomes simply  $\mu^z_{j-\frac{1}{2}}\tau^x_{j-1}\tau^x_{j}=1$. Thus the Rep$(D_{4n})$ IANISPT fixed-point state is the maximally mixed state in this constrained subspace:

\begin{widetext}
\[
\rho_{\text{IANISPT}}=\otimes_{j}\otimes^{n'-1}_{k=2, k\ne n/2+1}|\sigma^x_{n'(j-1)+k}=1\rangle\langle\sigma^x_{n'(j-1)+k}=1|\otimes \sum_{\{x_{j}\}}\bigotimes_{j}|\tau^x_{j}=x_{j}\rangle\langle \tau^x_{j}=x_{j}|\otimes
|\mu^z_{j-\frac{1}{2}}=x_{j-1}x_{j}\rangle\langle \mu^z_{j-\frac{1}{2}}=x_{j-1}x_{j}|.\nonumber
\]
\end{widetext}

\section{Rep($D_8$) IgSPT lattice model}
In this appendix, we generalize the gauging method to construct lattice models realizing Rep($D_8$) intrinsic gapless noninvertible SPT (igNISPT).

Similarly, we begin with $D_8$ side, where system is in the $\Z^U_2$ SSB phase with unbroken group $H=\Z^{\eta}_4$ and two ground states are both in $\Z^{\eta}_4$  igSPT phase. The Hamiltonian is given by
 \[\label{eq: SSB Hal app}
 \begin{split}
 H=&-\sum^L_{j=1}\sigma^z_{4j-3}\sigma^z_{4j+1}-
\sum^L_{j=1}\sigma^z_{4j-1}\sigma^z_{4j}\\
&-\frac{1}{2}\sum^L_{j=1}\sigma^z_{4j-2}\sigma^z_{4j-3}(\sigma^x_{4j-1}\sigma^x_{4j}\sigma^x_{4j-5}\sigma^x_{4j-4}\\&+\sigma^y_{4j-1}\sigma^y_{4j}\sigma^y_{4j-5}\sigma^y_{4j-4})
\\&-\sum^L_{j=1}\sigma^z_{4j}\sigma^z_{4j-3}(\sigma^x_{4j-2}\sigma^x_{4j+2}+\sigma^y_{4j-2}\sigma^y_{4j+2})
\\&-h\sum(\sigma^z_{4j-4}\sigma^z_{4j-3}+\sigma^z_{4j-3}\sigma^z_{4j-2}).
\end{split}
 \]
These five terms commute with each other. The first two terms in this Hamiltonian indeed impose the low energy Hilbert space is the subspace \eqref{eq: const-1} where $\sigma^z_{4j-3}$ is in the SSB phase. Then the other three terms becomes effectively:
\[
\begin{split}
H_{\text{eff}}=&-\sum^L_{j=1}\sigma^z_{4j-3}\mu^z_{j-\frac{1}{2}}\tau^x_{j-1}\tau^x_{j}\\&-\sum^L_{j=1}\sigma^z_{4j-3}\tau^z_j(\mu^x_{j-\frac{1}{2}}\mu^x_{j+\frac{1}{2}}+\mu^y_{j-\frac{1}{2}}\mu^y_{j+\frac{1}{2}})\\
&-h\sum^L_{j=1}\sigma^z_{4j-3}(\tau^z_{j-1}+\mu^z_{j-\frac{1}{2}}),
\end{split}
\]
where we use the notation of effective Pauli operators in \eqref{eq: eff Paul1}. In each SSB sector $\sigma^z_{4j-3}=\pm 1$, the system is described by a $\Z^{\eta}_4$ igSPT Hamiltonian with a magnetic field with strength $h$. The phase diagram of such Hamiltonian has been studied in Ref.~\cite{10.21468/SciPostPhys.18.5.153}:
\begin{enumerate}
    \item When $0\leq h <1$, each SSB sector belong to the intrinsic gapped SPT phase. 
    \item When $1<h<2$, each SSB sector realize a trivial gapless SPT phase.
    \item  When $h>2$: Only the magnetic field term  dominates in this regime, and hence each SSB sector is in the trivially gapped phase.
    \item When $h=1$, there is a phase transition in each SSB sector with center charge $c=3/2$ and when $h=2$, each SSB sector is in the Lifshitz transition.
\end{enumerate}

\begin{figure}[!tbp]
	\centering
	    \begin{tikzpicture}
	    \draw[thick,->] (0,0) -- (8.5,0);
	    \fill (3,0) circle (3pt);
	    \fill (6,0) circle (3pt);
            \fill (0,0) circle (3pt);
	      \node[below] at (0,0) {$h=0$};
	    \node[below] at (8.2,0) {$h= \infty$};
	    \node[below] at (3,0) {$h=1$};
	    \node[below] at (6,0) {$h=2$};
	      \draw[dashed, very thick] (3,0) -- (3,1.5);
            \draw[dashed, very thick] (6,0) -- (6,1.5);
	    \node at (1.2, 1) {\begin{tabular}{c}
	         Rep($D_8$) igNISPT
	    \end{tabular} };
	     \node at (4.5, 1) {\begin{tabular}{c}
	         Trivial   \\
	         Rep($D_8$)  gNISPT
	    \end{tabular} };
	   \node at (7.3, 1) {\begin{tabular}{c}
	         Trivially   \\
	         gapped Phase
	    \end{tabular} };

	    \node[above] at (3,-1.2) {\begin{tabular}{c}
	         ($c=3$)
	    \end{tabular}};
	   
	    \node[above] at (6,-1.2) {\begin{tabular}{c}
	         ($z=2$)
	    \end{tabular}};
	   
		\end{tikzpicture}
		\caption{The phase diagram of model \eqref{eq: dual Hal app}. }
		\label{fig:phaseigSPT}
\end{figure}
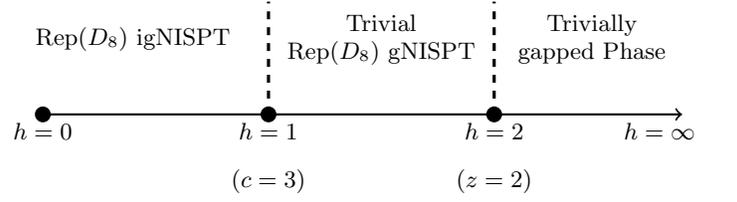

After KW duality, we obtain  the dual Hamiltonian
\[\label{eq: dual Hal app}
\begin{split}
 H=&- \sum^L_{j=1}\sigma^x_{4j-1}-  \sum^L_{j=1}\sigma^x_{4j-3}\sigma^x_{4j-2}\sigma^x_{4j}\\&-\frac{1}{2}\sum^L_{j=1}\sigma^z_{4j-6}\sigma^z_{4j-4}\sigma^x_{4j-3}\sigma^z_{4j-2}\sigma^z_{4j}(1+\sigma^x_{4j-5}\sigma^x_{4j-1})
 \\&+\sum^L_{j=1}(\sigma^y_{4j-3}\sigma^y_{4j-2}\sigma^x_{4j-1}\sigma^{z}_{4j+1}\sigma^{z}_{4j+2}\\&+\sigma^y_{4j-3}\sigma^z_{4j-2}\sigma^x_{4j}\sigma^y_{4j+1}\sigma^{z}_{4j+2})\\
 &-h\sum^L_{j=1}(\sigma^x_{4j}+\sigma^x_{4j+1}).
 \end{split}
\]
Similarly, all terms above commute with each other. The first two terms in the above Hamiltonian indeed impose the low energy Hilbert space is the subspace \eqref{eq:noninv subspace}. Thus the other terms becomes effectively in the low energy: 
\[
\begin{split}
H_{\text{eff}}=&-\sum^L_{j=1}\mu^z_{j-\frac{1}{2}}\tau^x_{j-1}\tau^x_{j}-\sum^L_{j=1}\tau^z_j(\mu^x_{j-\frac{1}{2}}\mu^x_{j+\frac{1}{2}}+\mu^y_{j-\frac{1}{2}}\mu^y_{j+\frac{1}{2}})\\
&-\sum^L_{j=1}(\mu^x_{j-\frac{1}{2}}+\tau^x_{j}),
\end{split}
\]
where we use the notation of effective Pauli operators in \eqref{eq: eff Paul2}. In the low energy Hilbert space, the effective symmetry is given by \eqref{eq: eff sym} and the effective Hamiltonian is indeed the $\Z^{\eta}_4$ igSPT Hamiltonian with a magnetic field with strength $h$. By applying KW duality on the phase and phase transition of the system \eqref{eq: SSB Hal app}, we obtain the phase diagram of the dual system \eqref{eq: dual Hal app}  as shown in  Fig.~\ref{fig:phaseigSPT}.

\begin{enumerate}
    \item When $0\leq h <1$, the system belongs to the Rep($D_8$) igNISPT phase. 
    \item When $1<h<2$, the system belongs to a trivial gapless NISPT (gNISPT) phase.
    \item  When $h>2$: The magnetic field term  dominates and hence the system is in the trivially gapped phase.
    \item When $h=1$, the phase transition has the center charge $c=3$ and when $h=2$, the phase transition is the Lifshitz transition.
\end{enumerate}

\end{document}

%% file: duality.tex
\begin{tikzpicture}
\draw[red] (-3, 0.25) -- (7, 0.25) -- (7, -1+0.25) -- (-3, -1+0.25) --(-3, 0.25);
    \begin{scope}[shift={(0, 0)}] 
    \node [] at (-3+0.3+0.5,-0.25) {\small Key idea:};
    \node [] at (-3+0.3+0.5,-1.5) {\small Case 1:};
     \node [] at (-3+0.3+0.5,-3) {\small Case 2:};
    \node [] at (0,0) {\small Mixed state phase };
    \node [] at (0,-0.5) {\small with symmetry $\mathcal C_1$};
    \node [] at (0,-1.5+0.25) {\small $\mathcal C_1$=Rep$(D_{2n})$};
    \node [] at (0,-1.5-0.25) {\small ASPT};
        \node [] at (0,-3+0.25) {\small \small $\mathcal C_1$=$(\mathbb Z_N^Q\times\mathbb Z_N^D)\rtimes T$};
    \node [] at (0,-3-0.25) {\small ASPT};
    \end{scope}

    \begin{scope}[shift={(5.5, 0)}] 
    \node [] at (0,0) {\small Mixed state phase};
    \node [] at (0,-0.5) {\small with symmetry $\mathcal C_2$};
    \node [] at (0,-1.5+0.25) {\small \small $\mathcal C_2$=$D_{2n}=\mathbb Z_n^{\eta}\rtimes \mathbb Z_2^U$};
    \node [] at (0,-1.5-0.25) {\small $\mathbb Z_2^{U}$ SSB}; 
     \node [] at (0,-3+0.25) {\small \small $\mathcal C_2$=$(\mathbb Z_N^Z\times\mathbb Z_N^X)\rtimes T$};
    \node [] at (0,-3-0.25) {$\mathbb Z_N^X$ SSB}; 
    \end{scope}

    \draw[<->, thick] (3-1.25,-0.25) -- (5-1.25, -0.25) node[midway, above] {\footnotesize Gauging};
\draw[<->, thick] (3-1.25,-1.5) -- (5-1.25, -1.5) node[midway, above] {\footnotesize Gauging $\mathbb Z_2$ };
\draw[<->, thick] (3-1.25,-3) -- (5-1.25, -3) node[midway, above] {\footnotesize Gauging $\mathbb Z_N$};
 
\end{tikzpicture}